\documentclass[structabstract]{aa}%[traditabstract]{aa}  %referee,
\usepackage{longtable}
\usepackage{amssymb,times,graphicx, subfigure}
\usepackage[english]{babel}
\usepackage{txfonts}
\usepackage{xspace}
\usepackage{multirow}
\usepackage[usenames]{color}
\usepackage[normalem]{ulem}
\usepackage{natbib,afterpage}
\bibpunct{(}{)}{;}{a}{,}{,}
\def\irc{IRC\,+10216\xspace}
\def\um{$\mu$m}
\def\ethynyl{C$_2$H\xspace}
\def\acetylene{C$_2$H$_2$\xspace}

\def\nutwo{$\nu_2$\xspace}

\def\onesigma{$^1\Sigma$\xspace}

\def\twosigmaplus{$^2\Sigma^+$\xspace}

\def\bond#1{\relax
    \hbox \bgroup
       \kern 0.25em
       \setbox0=\hbox{X}%
       \vbox to \ht0 \bgroup
          \vss
          \ifcase#1
          \or
             \hrule width 0.7em depth 0pt height 0.03em
          \or
             \hrule width 0.7em depth 0pt height 0.03em
             \vskip 0.2em%changed jms89 from 0.1
             \hrule width 0.7em depth 0pt height 0.03em
          \or
             \hrule width 0.7em depth 0pt height 0.03em
             \vskip 0.2em%changed jms89 from 0.1
             \hrule width 0.7em depth 0pt height 0.03em
             \vskip 0.2em%changed jms89 from 0.1
             \hrule width 0.7em depth 0pt height 0.03em
          \fi
          \vss
          \egroup
       \kern 0.25em
       \egroup
    }%

\begin{document}
\title{On the physical structure of IRC\,+10216}
\subtitle{Ground-based and \emph{Herschel}\thanks{\emph{Herschel} is an ESA space observatory with science instruments provided by European-led Principal Investigator consortia and with important participation from NASA.} observations of CO and \ethynyl}
\titlerunning{CO and CCH around IRC\,+10216}
\author{  E. De Beck\inst{1}
	  \and R. Lombaert\inst{1}
	  \and M. Ag\'undez\inst{2,3}
	  \and F. Daniel\inst{2}
	  \and L. Decin\inst{1,4}
	  \and J. Cernicharo\inst{2}
	  \and H. S. P. M\"uller\inst{5}
	  \and M. Min\inst{6}
	  \and P. Royer\inst{1}
	  \and B. Vandenbussche\inst{1}
	  \and A. de Koter\inst{4,6}
	  \and L. B. F. M. Waters\inst{7,4}	  
	  \and M. A. T. Groenewegen\inst{8} % for MESS
	  \and M. J. Barlow\inst{9} % for MESS
	  \and M. Gu{\'e}lin\inst{10,13}
	  \and C. Kahane\inst{11}
	  \and J. C. Pearson\inst{12}
	  \and P. Encrenaz\inst{13}
	  \and R. Szczerba\inst{14}
	  \and M. R. Schmidt\inst{14}
          }
\institute{Institute for Astronomy, Department of Physics and Astronomy, KULeuven, Celestijnenlaan 200D, 3001 Heverlee, Belgium\\
              \email{elvire.debeck@ster.kuleuven.be}
             \and CAB. INTA-CSIC. Crta Torrej\'on km 4. 28850 Torrej\'on de Ardoz. Madrid. Spain
	     	 \and LUTH, Observatoire de Paris-Meudon, 5 Place Jules Janssen, 92190 Meudon, France
             \and Astronomical Institute ``Anton Pannekoek'', University of Amsterdam, Science Park 904, 1098 XH Amsterdam, The Netherlands
	    	 \and I. Physikalisches Institut, Universit\"at zu K\"oln, Z\"ulpicher Str. 77, 50937 K\"oln, Germany 
	    	 \and Astronomical Institute Utrecht, University of Utrecht, PO Box 8000, NL-3508 TA Utrecht, The Netherlands
	    	 \and SRON Netherlands Institute for Space Research, Sorbonnelaan 2, 3584 CA Utrecht, The Netherlands
	    	 \and Royal Observatory of Belgium, Ringlaan 3, 1180 Brussels, Belgium
	    	  \and Department of Physics and Astronomy, University College London, Gower Street, London WC1E 6BT, UK  
	     	 \and Institut de Radioastronomie Millim\'etrique (IRAM), 300 rue de la Piscine, 38406 Saint-Martin-d’H\`eres, France 
	     	 \and LAOG, Observatoire de Grenoble, UMR 5571-CNRS, Universit\'e Joseph Fourier, Grenoble, France 
	     	 \and Jet Propulsion Laboratory, Caltech, Pasadena, CA 91109, USA 
	     	 \and LERMA, CNRS UMR8112, Observatoire de Paris and {\'E}cole Normale Sup{\'e}rieure, 24 Rue Lhomond, 75231 Paris Cedex 05, France 
	     	 \and N. Copernicus Astronomical Center, Rabianska 8, 87-100 Torun, Poland
}

\date{Received ---; accepted ---}

\abstract
{
The carbon-rich asymptotic giant branch star \irc undergoes strong mass loss, and quasi-periodic enhancements of the density of the circumstellar matter have previously been reported. The star's circumstellar environment is a well-studied, and complex astrochemical laboratory, with many molecular species proved to be present. CO is ubiquitous in the circumstellar envelope, while emission from the ethynyl (\ethynyl) radical is detected in a spatially confined shell around \irc. As reported in this article, we recently detected unexpectedly strong emission from the $N=4-3,\,6-5,\,7-6,\,8-7$, and $9-8$ transitions of \ethynyl with the IRAM 30\,m telescope and with \emph{Herschel}/HIFI, challenging the available chemical and physical models.
}
{
We aim to constrain the physical properties of the circumstellar envelope of \irc, including the effect of episodic mass loss on the observed emission lines. In particular, we aim to determine the excitation region and conditions of \ethynyl, in order to explain the recent detections, and to reconcile these with interferometric maps of the $N=1-0$ transition of \ethynyl.
}
{
Using radiative-transfer modelling, we provide a physical description of the circumstellar envelope of \irc, constrained by the spectral-energy distribution and a sample of 20 high-resolution and 29 low-resolution CO lines --- to date, the largest modelled range of CO lines towards an evolved star. We further present the most detailed radiative-transfer analysis of \ethynyl that has been done so far.
}
{Assuming a distance of 150\,pc to \irc, the spectral-energy distribution is modelled with a stellar luminosity of 11300\,$L_{\sun}$ and a dust-mass-loss rate of $4.0\times10^{-8}\,M_{\sun}$\,yr$^{-1}$. Based on the analysis of the 20 high-frequency-resolution CO observations, an average gas-mass-loss rate for the last 1000\,years of $1.5\times10^{-5}$\,$M_{\sun}$\,yr$^{-1}$ is derived. This results in a gas-to-dust-mass ratio of 375, typical for this type of star. The kinetic temperature throughout the circumstellar envelope is characterised by three powerlaws: $T_{\mathrm{kin}}(r)\propto r^{-0.58}$ for radii $r\leq 9$ stellar radii, $T_{\mathrm{kin}}(r)\propto r^{-0.40}$ for radii $9\leq r\leq 65$ stellar radii, and $T_{\mathrm{kin}}(r)\propto r^{-1.20}$ for radii $r\geq 65$ stellar radii. This model successfully describes all 49 observed CO lines. %The presented model successfully describes the excitation of all 49 observed CO lines.
We also  show the effect of density enhancements in the wind of \irc on the \ethynyl-abundance profile, and the close agreement we find of the model predictions with interferometric maps of the \ethynyl $N=1-0$ transition and with the rotational lines observed with the IRAM 30\,m telescope and \emph{Herschel}/HIFI. We report on the importance of radiative pumping to the vibrationally excited levels of \ethynyl and the significant effect this pumping mechanism has on the excitation of all levels of the \ethynyl-molecule. 
}
{}


\maketitle

% ###############################################################################
% ###############################################################################
% ###############################################################################

\section{Introduction}\label{sect:intro}
The carbon-rich Mira-type star \irc (CW\,Leo) is located at the tip of the asymptotic giant branch (AGB), where it loses mass at a high rate \citep[$\sim$$1-4\times10^{-5}$\,$M_{\sun}$\,yr$^{-1}$;][]{crosas1997,groenewegen1998,cernicharo2000_2mm,debeck2010}.  Located at a distance of $120-250$\,pc  \citep{loup1993,crosas1997,groenewegen1998,cernicharo2000_2mm}, it is the most nearby C-type AGB star. Additionaly, since its very dense circumstellar envelope (CSE) harbours a rich molecular chemistry, it has been deemed the prime carbon-rich AGB astrochemical laboratory. More than 70 molecular species have already been detected \citep[e.g. ][]{cernicharo2000_2mm,he2008,tenenbaum2010}, of which many are carbon chains, e.g. cyanopolyynes HC$_n$N ($n$=1, 3, 5, 7, 9, 11) and C$_n$N ($n$=1, 3, 5). Furthermore, several anions have been identified, e.g. C$_n$H$^-$  \citep[$n=4,6,8$;][]{cernicharo2007,remijan2007,kawaguchi2007}, and C$_3$N$^-$ \citep{thaddeus2008}, C$_5$N$^-$ \citep{cernicharo2008}, and CN$^-$ \citep{agundez2010_cnmin}. Detections towards \irc of acetylenic chain radicals (C$_n$H), for $n$=2 up to $n$=8, have been reported by e.g. \citet[C$_4$H]{guelin1978_c4hdetection}, \citet[C$_5$H]{cernicharo1986_c5hdetection,cernicharo1986_c5htentativedetection,cernicharo1987_c5hhyperfine}, \cite{cernicharo1987_c6hhyperfine} and \citet[C$_6$H]{guelin1987_c6hdetection}, \citet[C$_7$H]{guelin1997_c7h}, and \citet[C$_8$H]{cernicharo1996_c8hdetection}.

The smallest C$_n$H radical, ethynyl (CCH, or \ethynyl), was first detected by \cite{tucker1974} in its $N=1-0$ transition in the interstellar medium (ISM) and in the envelope around \irc. It was shown to be one of the most abundant ISM molecules. The formation of \ethynyl in the envelope of \irc is attributed mainly to photodissociation of \acetylene (acetylene), one of the most prominent molecules in carbon-rich AGB stars. \cite{fonfria2008} modelled \acetylene emission in the mid-infrared, which samples the dust-formation region in the inner CSE. The lack of a permanent dipole moment in the linearly symmetric \acetylene-molecule implies the absence of pure rotational transitions which typically trace the outer parts of the envelope. \ethynyl, on the other hand, has prominent rotational lines that probe the chemical and physical conditions linked to \acetylene  in these cold outer layers of the CSE. It has been established that \ethynyl emission arises from a shell of radicals situated at $\sim$15\arcsec\xspace from the central star \citep{guelin1993}. Observations with the IRAM 30\,m telescope and \emph{Herschel}/HIFI show strong emission in several high-$N$ rotational transitions of \ethynyl, something which is unexpected and challenges our understanding of this molecule in \irc.

We present and discuss the high-sensitivity, high-resolution data obtained with the instruments on board \emph{Herschel} \citep{pilbratt2010}: HIFI \citep[Heterodyne Instrument for the Far Infrared;][]{degraauw2010}, SPIRE \citep[Spectral and Photometric Imaging Receiver;][]{griffin2010}, and PACS \citep[Photodetector Array Camera \& Spectrometer;][]{poglitsch2010} and with the IRAM 30\,m Telescope\footnote{Based on observations carried out with the IRAM 30\,m Telescope. IRAM is supported by INSU/CNRS (France), MPG (Germany) and IGN (Spain).} in Sect.~\ref{sect:observations}.  The physical model for \irc's circumstellar envelope is presented and discussed in Section~\ref{sect:model}. The \ethynyl molecule, and our treatment of it is described in Section~\ref{sect:cch}. A summary of our findings is provided in Section~\ref{sect:conclusions}.

\begin{figure*}[ht]\centering
\includegraphics[angle=90,width=\linewidth]{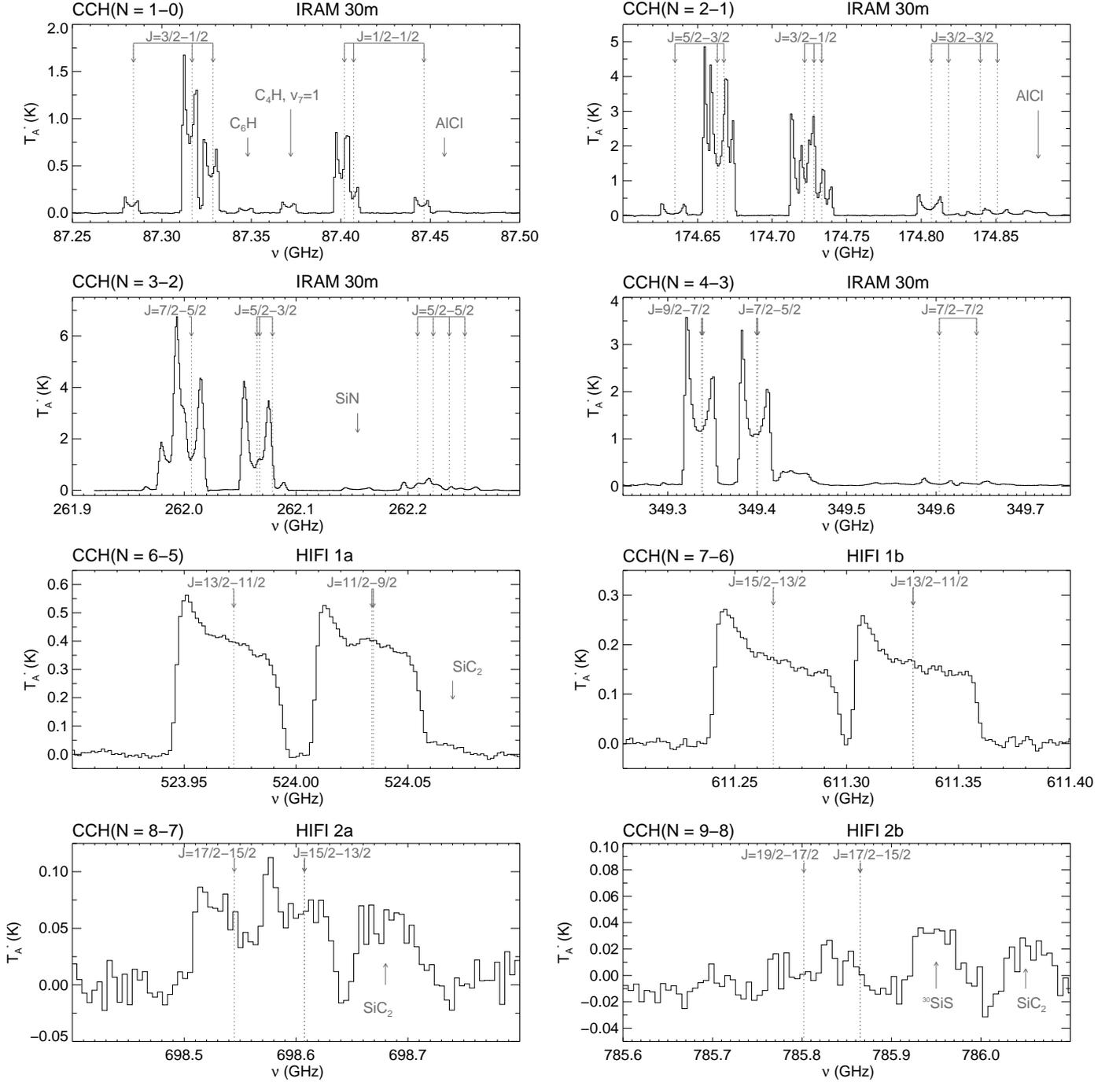}
\caption{\ethynyl $N=1-0$, $2-1$, $3-2$, $4-3$, $6-5$, $7-6$, $8-7$, and $9-8$ rotational transitions in \irc's envelope as observed with the IRAM 30\,m telescope and \emph{Herschel}/HIFI. The fine structure components are labelled with the respective $J$-transitions in the top of each panel, while the hyperfine components are indicated with vertical dotted lines.
%The hyperfine structure components of the fine structure are labelled with vertical dotted lines.
%The fine structure, i.e. the $J$ transitions, is labelled with horizontal grey lines in the plots, grouping the appropriate hyperfine structure components,  which are indicated with vertical dotted lines. 
For the sake of clarity, we omitted the components that are too weak to be detected. See Sect.~\ref{sect:cch} for details on the spectroscopic structure of \ethynyl. Additionally observed features related to other molecules are also identified.} \label{fig:cch_all} 
\end{figure*}

\begin{table*}[ht]\caption{Summary of the IRAM 30\,m and HIFI detections of \ethynyl. Columns are, respectively, the instrument or band with which we detected the \ethynyl emission, the main-beam efficiency $\eta_{\mathrm{MB}}$, the half-power beam width HPBW, the integration time, the noise level in the spectra, the frequency resolution $\Delta\nu$, the velocity resolution $\Delta\varv$, the detected \ethynyl rotational transitions, the frequency range in which we observed the lines, and the frequency-integrated intensity $I_{\nu,\mathrm{A}}^{*}=\int{T_\mathrm{A}^{*}d\nu}$. }\label{tbl:iramhifi}
\begin{center}
\setlength{\tabcolsep}{2.3mm}
  \begin{tabular}{cccccc%c
cccc}
\hline\hline\\[-2ex]
Instrument	&$\eta_{\mathrm{MB}}$	&HPBW&Int.Time	& Noise	&$\Delta\nu$ 	&$\Delta\varv$& CCH & Freq. Range&$I_{\nu,\mathrm{A}}^{*}$\\
or Band &&(\arcsec)& (min)		& (mK)	&(MHz)		&(km\,s$^{-1}$)& transition 	& (MHz) 				&(K\,MHz)\\
\hline\\[-2ex]
\textbf{IRAM 30\,m}\\
A100/B100	&0.82	&28.2&614	&2.5	&1.0&3.4	&$N=1-0$& \\
	&&&&&&&$J=3/2-1/2$&$87278.2-87339.6$&15.90\\
	&&&&&&&$J=1/2-1/2$&$87392.8-87453.6$&7.58\\
C150/D150  	&0.68	&14.1&215	&10	&1.0	&1.7	&$N=2-1$& \\
	&&&&&&&$J=5/2-3/2$&$174621.1-174682.5$&58.73\\
	&&&&&&&$J=3/2-1/2$&$174708.2-174744.9$&38.09\\
	&&&&&&&$J=3/2-3/2$&$174796.2-174863.7$&5.90\\
E3 	&0.57&9.4&161	&10	&1.0	&1.1&$N=3-2$& \\
	&&&&&&&$J=7/2-5/2$&$261974.3-262023.4$&110.74\\
	&&&&&&&$J=5/2-3/2$&$262048.8-262084.4$&60.63\\
	&&&&&&&$J=5/2-5/2$&$262191.3-262270.0$&6.33\\
E3 	&0.35&7.0&79	&	5.3&2.0	&1.7&$N=4-3$& \\
	&&&&&&&$J=9/2-7/2$ &$349311.5-349368.3$&66.98\\
	&&&&&&&$J=7/2-5/2$ &$349368.3-349421.6$&60.15\\
	&&&&&&&$J=7/2-7/2$&$349575.1-349671.8$&2.35\\
\hline\\[-2ex]
\textbf{\emph{Herschel}/HIFI}\\

1a	&0.75&40.5& 28&9.9	&1.5	&0.86	&$N=6-5$& \\
	&&&&&&&$J=13/2-11/2$&$523941.9-524000.4$&18.60\\
	&&&&&&&$J=11/2-9/2$ &$524000.4-524061.9$&18.74\\
	&&&&&&&$J=11/2-11/2$&$-$&$-$\\
1b	&0.75&34.7 & 13&7.4 	&0.50	&0.25	&$N=7-6$&\\
	&&&&&&&$J=15/2-13/2$ &$611237.1-611301.6$&10.09\\
	&&&&&&&$J=13/2-11/2$ &$611301.6-611364.1$&9.44\\
	&&&&&&&$J=13/2-13/2$ &$-$&$-$\\
2a\tablefootmark{(\dagger)}&0.75&30.4&12	&13.7	&4.5&1.8	&$N=8-7$&\\
	&&&&&&&$J=17/2-15/2$&$698487.5-698619.0$&4.21\\
	&&&&&&&$J=15/2-13/2$&$698619.0-698738.0$&4.47\\
	&&&&&&&$J=15/2-15/2$&$-$&$-$\\
2b&0.75&27.0&15&10	&6.0	&2.2&$N=9-8$&\\
	&&&&&&&$J=19/2-17/2$&$785750.0-785875.0$\tablefootmark{(\ddagger)}&0.20\tablefootmark{(\ddagger)}\\
	&&&&&&&$J=17/2-15/2$&$785875.0-785900.0$\tablefootmark{(\ddagger)}&0.20\tablefootmark{(\ddagger)}\\
	&&&&&&&$J=17/2-17/2$&$-$&$-$\\
\hline\\ [-6ex]
 \end{tabular}
\end{center}
\tablefoot{
\tablefoottext{\dagger}{Only the horizontal polarisation was obtained.}
\tablefoottext{\ddagger}{Due to the low signal-to-noise ratio of this data set, these numbers are to be interpreted with caution.}
}
\end{table*}

\begin{figure*}
\centering
\includegraphics[angle=90,width=\linewidth]{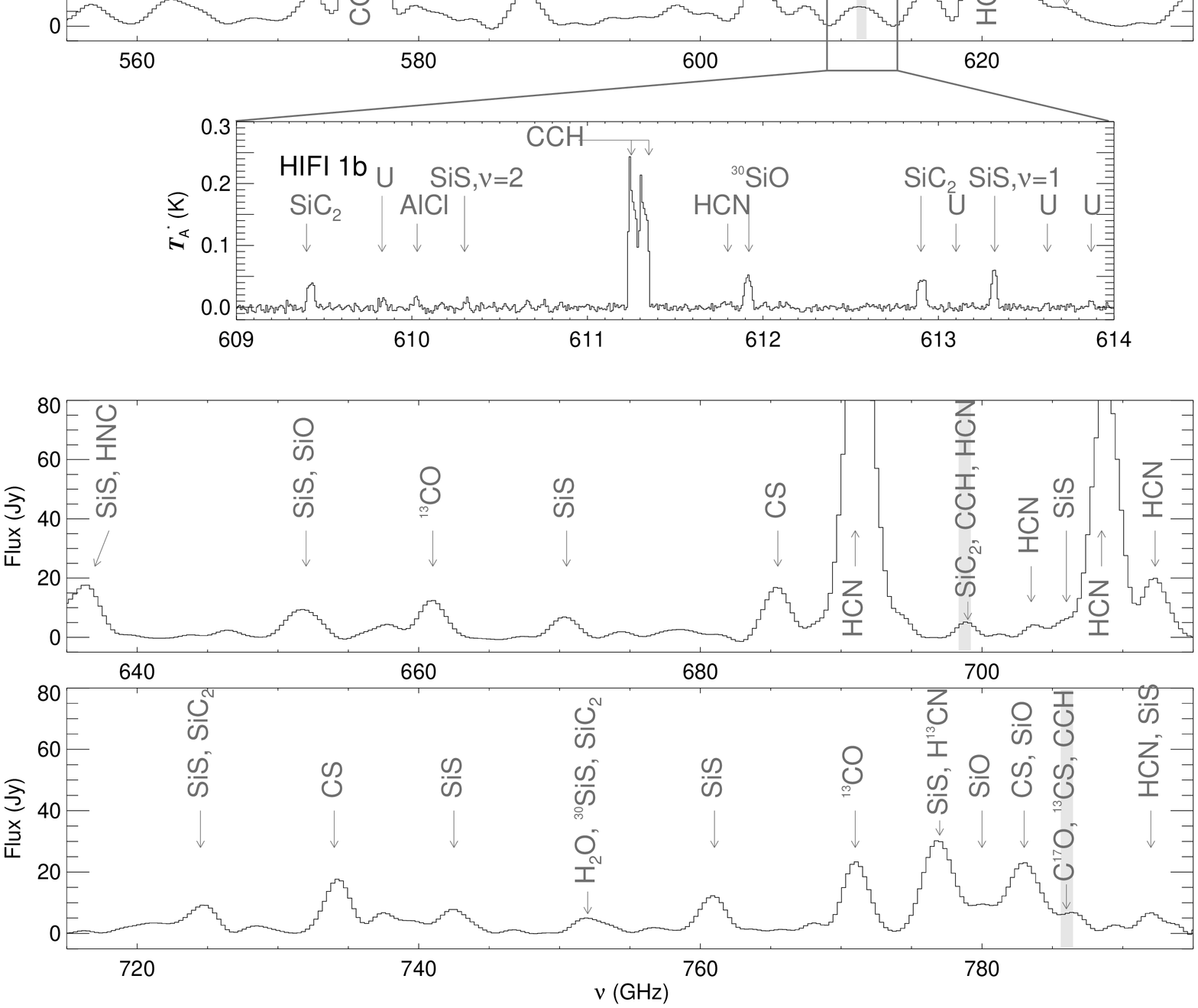}
\vspace{.5cm}
\caption{SPIRE data in the range $475-795$\,GHz, containing the \ethynyl-lines detected with HIFI. The shaded strips in the panels indicate the spectral ranges of the HIFI data shown in the panels of Fig.~\ref{fig:cch_all}. The labels mark the principal contributors to the main spectral features. A comparison of emission in the range $609-614$\,GHz, as obtained with SPIRE and with HIFI, is shown in the inset panel. \label{fig:spire_hifi}}
\end{figure*}

\section{Observations}\label{sect:observations}
\subsection{Herschel/HIFI: Spectral Scan}\label{sect:hifi}
The HIFI data of \irc presented in this paper are part of a spectral line survey carried out with HIFI's wideband spectrometer \citep[WBS; ][]{degraauw2010} in May 2010, on 3 consecutive operational days (ODs) of the \emph{Herschel} mission. The scan covers the frequency ranges $480-1250$\,GHz and $1410-1910$\,GHz, with a spectral resolution of 1.1\,MHz. All spectra were measured in dual beam switch \citep[DBS; ][]{degraauw2010} mode with a 3\arcmin\xspace chop throw. This technique allows one to correct for any off-source signals in the spectrum, and to obtain a stable baseline. 

Since HIFI is a double sideband (DSB) heterodyne instrument, the measured spectra contain lines pertaining to both the upper and the lower sideband. The observation and data-reduction strategies disentangle these sidebands with very high accuracy, producing a final single-sideband (SSB) spectrum without ripples, ghost features, and any other instrumental effects, covering the spectral ranges mentioned above.

The two orthogonal receivers of HIFI (horizontal H, and vertical V) were used simultaneously to acquire data for the whole spectral scan, except for band 2a\footnote{Because of an incompleteness, only the horizontal receiver was used during observations in band 2a. Supplementary observations will be executed later in the mission.}. Since we do not aim to study polarisation of the emission, we averaged the spectra from both polarisations, reducing the noise in the final product. This approach is justified since no significant differences between the H and V spectra are seen for the lines under study.

A detailed description of the observations, and of the data reduction of this large survey is given by \cite{cernicharo2010_hifi1b} and \citet[\emph{in prep.}]{cernicharo2011_scan}. All \ethynyl data in this paper are presented in the antenna temperature ($T_\mathrm{A}^{*}$) scale\footnote{The intensity in main-beam temperature $T_{\mathrm{MB}}$ is obtained via $T_{\mathrm{MB}}=T_\mathrm{A}^{*}/\eta_{\mathrm{MB}}$, where $\eta_{\mathrm{MB}}$ is the main-beam efficiency, listed in Table~\ref{tbl:iramhifi}.}. \citet[\emph{submitted}]{roelfsema2012} describe the calibration of the instrument and mention uncertainties in the intensity of the order of 10\%.

For the current study we concentrate on observations of CO and \ethynyl. The ten CO transitions covered in the survey range from $J=5-4$ up to $J=11-10$ and from $J=14-13$ up to $J=16-15$. The \ethynyl rotational transitions $N=6-5$, $7-6$, and $8-7$ are covered by the surveys in bands 1a ($480-560$\,GHz), 1b ($560-640$\,GHz), and 2a ($640-720$\,GHz), respectively. The $N=9-8$ transition is detected in band 2b ($720-800$\,GHz) of the survey, but with very low S/N; higher-$N$ transitions were not detected. A summary of the presented HIFI data of \ethynyl is given in Table~\ref{tbl:iramhifi}, the spectra are shown in Fig.~\ref{fig:cch_all}.

\subsection{Herschel/SPIRE}\label{sect:spire}
In the framework of the Herschel guaranteed time key programme ``Mass loss of Evolved StarS'' \citep[MESS; ][]{groenewegen2011} the SPIRE Fourier Transform Spectrometer \citep[FTS; ][]{griffin2010} was used to obtain \irc's spectrum on 19 November 2009 (OD 189). The SPIRE FTS measures the Fourier transform of the source spectrum across two wavelength bands simultaneously. The short wavelength band (SSW) covers the range $194-313\,\mu$m, while the long wavelength band (SLW) covers the range $303-671\,\mu$m. The total spectrum covers the $446-1575$\,GHz frequency range, with a final spectral resolution of 2.1\,GHz. The quality of the acquired data permits the detection of lines as weak as $1-2$\,Jy. For the technical background and a description of the data-reduction process we refer to the discussions by \cite{cernicharo2010_hcl} and \cite{decin2010_silicon}. These authors report uncertainties on the SPIRE FTS absolute fluxes of the order of $15-20$\% for SSW data, $20-30$\% for SLW data below 500\,$\mu$m, and up to 50\% for SLW data beyond 500\,$\mu$m. We refer to Table~\ref{tbl:spire-obs} for a summary of the presented data, and show an instructive comparison between the HIFI and SPIRE data of the \ethynyl lines $N=6-5$ up to $N=9-8$ in Fig.~\ref{fig:spire_hifi}. It is clear from these plots that the lower resolution of the SPIRE spectrum causes many line blends in the spectra of AGB stars.

\subsection{Herschel/PACS}\label{sect:pacs}
\cite{decin2010_silicon} presented PACS data of \irc, also obtained in the framework of the MESS programme \citep{groenewegen2011}. The full data set consists of SED scans in the wavelength range $52-220$\,$\mu$m, obtained at different spatial pointings on November 12, 2009 (OD\,182). We re-reduced this data set, taking into account not only the central spaxels of the detector, but all spaxels containing a contribution to the flux, i.e. 20 out of 25 spaxels in total. The here presented data set, therefore, reflects the total flux emitted by the observed regions, within the approximation that there is no loss between the spaxels. The estimated uncertainty on the line fluxes is of the order of 30\%. Due to instrumental effects, only the range $56-190$\,$\mu$m of the PACS spectrum is usable. This range holds $^{12}$CO transitions $J=14-13$ up to $J=42-41$, covering energy levels from $\sim$350\,cm$^{-1}$ up to $\sim$3450\,cm$^{-1}$. As is the case for the SPIRE data, line blends are present in the PACS spectrum due to the low spectral resolution of $0.08-0.7$\,GHz.

\subsection{IRAM 30\,m: line surveys} \label{sect:iram}
We combine the \emph{Herschel} data with data obtained with the IRAM 30\,m telescope at Pico Veleta. The \ethynyl $N=1-0$ transition was observed by \cite{kahane1988}; $N=2-1$ was observed by \cite{cernicharo2000_2mm}. $N=3-2$ and $N=4-3$ were observed between January and April 2010, using the EMIR receivers as described in detail by \citep[][\emph{in prep.}]{kahane2011_1mm}. The data-reduction process of these different data sets is described in detail in the listed papers. The observed \ethynyl transitions are shown in Fig.~\ref{fig:cch_all}, and details on the observations are listed in Table~\ref{tbl:iramhifi}.

\begin{table}[ht]\caption{Summarised information on the SPIRE spectrum containing the $N=7-6$ transition of \ethynyl. $\Delta\nu$ is the frequency resolution, $\Delta\varv$ is the velocity resolution, and $\sigma$ is the rms noise. $F_{\nu,\rm{A}^{*}}$ is the frequency-integrated flux in the given frequency range.}\label{tbl:spire-obs}
\centering
\begin{tabular}{cr|cr}
\hline\hline\\[-2ex]
Frequency range	&$446-1575$\,GHz &Integration time &2664\,s\\
$\Delta\nu$	&2.1\,GHz &$\Delta\varv$	&201\,km\,s$^{-1}$\\
$F_{\nu,\rm{A}^{*}}$&16915\,Jy\,MHz & $\sigma$&	850\,mJy\\
\hline
\end{tabular}
\end{table}

\section{The envelope model: dust \& CO}\label{sect:model}
We assumed a distance $d=150$\,pc to \irc, in good correspondence with literature values \citep[][and references therein]{groenewegen1998,menshchikov2001,schoeier2007}. The second assumption in our models is that the effective temperature $T_\mathrm{eff}= 2330$\,K, following the model of a large set of mid-IR lines of \acetylene and HCN, presented by \cite{fonfria2008}. We determined the luminosity $L_{\star}$, the dust-mass-loss rate $\dot{M}_\mathrm{dust}$, and the dust composition from a fit to the spectral energy distribution (SED; Sect.~\ref{sect:radtrandust}). The kinetic temperature profile and the gas-mass-loss rate $\dot{M}$ were determined from a CO-line emission model (Sect.~\ref{sect:radtrangas}). The obtained envelope model will serve as the basis for the \ethynyl-modelling presented in Sect.~\ref{sect:cch}. We point out that all modelling is performed in the radial dimension only, i.e. in 1D, assuming spherical symmetry throughout the CSE.

\subsection{Dust} \label{sect:radtrandust}
\begin{figure}
\includegraphics[width=.925\linewidth]{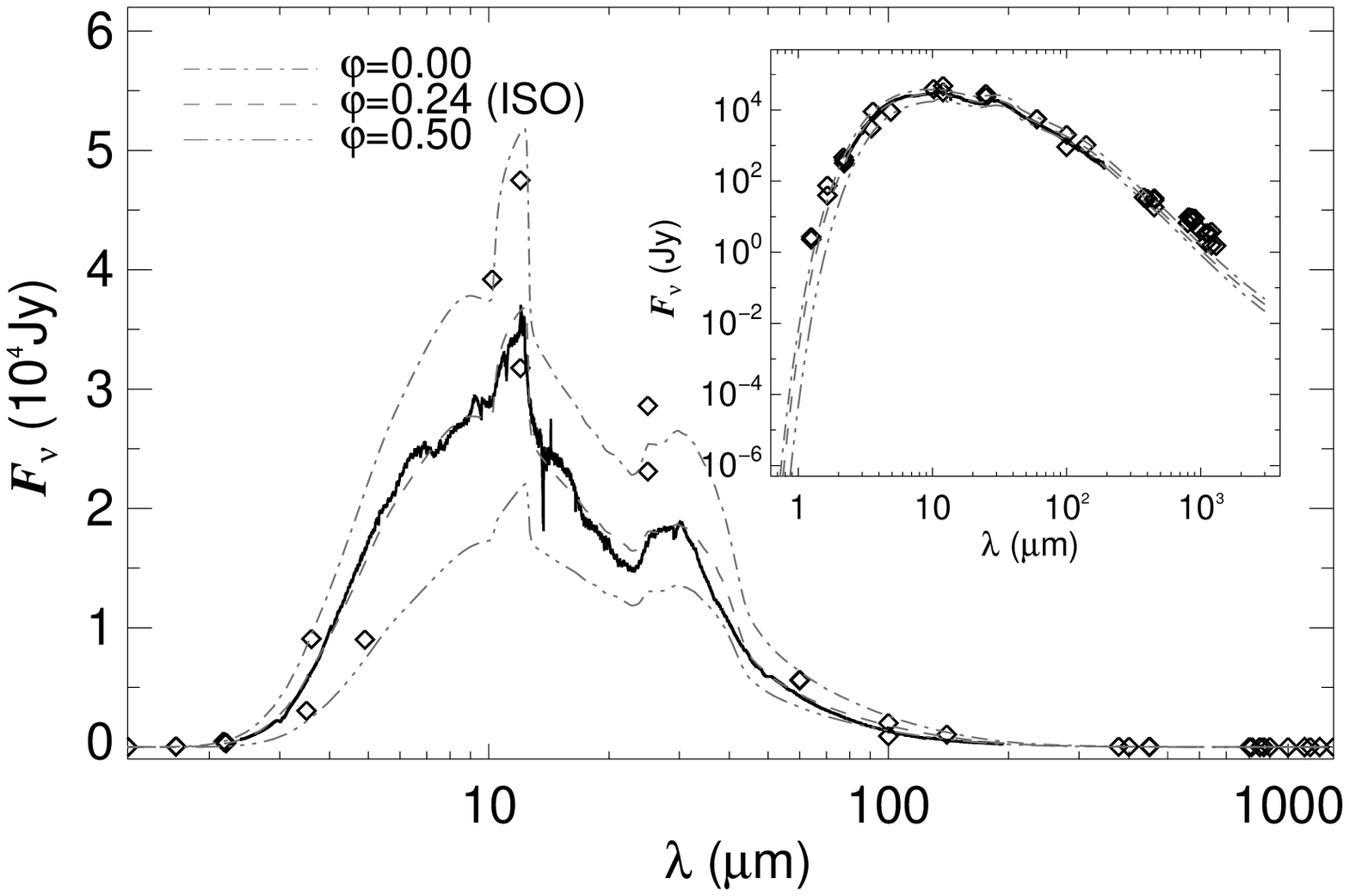}
\caption{SED models for \irc, using $\dot{M}_\mathrm{dust}=4.0\times10^{-8}$\,$M_{\sun}$\,yr$^{-1}$, and the dust composition and properties listed in Table~\ref{tbl:dustcomposition} and spatial distribution and properties shown in Fig.~\ref{fig:dustprops}. The SED model is constrained by the ISO SWS and LWS data \emph{(full black)}; photometric points \emph{(black diamonds)} are taken from \cite{ramstedt2008} and \cite{ladjal2010}. \emph{Dash-dotted grey:} model at maximum light, with 15800\,$L_{\sun}$, \emph{dashed grey:} best fit to the ISO data, at phase $\varphi$=0.24, using 11300\,$L_{\sun}$, \emph{dash-triple-dotted grey:} model at minimum light, with 6250\,$L_{\sun}$. The inset is identical, but plotted in $\log-\log$ scale.}\label{fig:sed}
\end{figure}

The dust modelling was performed using \texttt{MCMax} \citep{min2009_mcmax}, a Monte Carlo dust radiative transfer code. The best SED fit to the ISO SWS and LWS data, shown in Fig.~\ref{fig:sed}, is based on a stellar luminosity $L_{\star}$=11300\,$L_{\sun}$. This is in good correspondence with the results of \cite{menshchikov2001}\footnote{The extensive modelling of \citeauthor{menshchikov2001} was based on a light-curve analysis and SED modelling, and holds a quoted uncertainty on the luminosity of 20\%.}, considering the difference in adopted distance. The combination of $L_{\star}$ and $T_{\mathrm{eff}}$ gives a stellar radius $R_{\star}$ of 20\,milli-arcseconds, in very good agreement with the values reported by e.g. \citet[][19\,mas]{ridgway1988} and \citet[][22\,mas]{monnier2000}. 

\irc is a Mira-type pulsator, with a period of 649\,days \citep{lebertre1992}. Following Eq.~1  of \citeauthor{menshchikov2001}, we find that $L_{\star}$ varies between $L_{\star,\varphi=0}$$\approx$15800\,$L_{\sun}$ at maximum light (at phase $\varphi$=0), and $L_{\star,\varphi=0.5}$$\approx$6250\,$L_{\sun}$ at minimum light ($\varphi$=0.5). Fig.~\ref{fig:sed} shows the SED-variability corresponding to the $L_{\star}$-variability. From the figure, it is visible that the spread on the photometric points can be accounted for by the models covering the full $L_{\star}$-range.

The adopted dust composition is given in Table~\ref{tbl:dustcomposition}. The main constituents are amorphous carbon (aC), silicon carbide (SiC), and magnesium sulfide (MgS), with mass fractions of 53\%, 25\%, and 22\%, respectively. The aC grains are assumed to follow a distribution of hollow spheres \citep[DHS; ][]{min2003_dhs}, with size 0.01\,$\mu$m and a filling factor of 0.8. The population of the SiC and MgS dust grains is represented by a continuous distribution of ellipsoids \citep[CDE; ][]{bohren1983}, where the ellipsoids all have the volume of a sphere with radius $a_{\mathrm{d}}$=0.1\,$\mu$m. CDE and DHS are believed to give a more realistic approximation of the characteristics of circumstellar dust grains than a population of spherical grains (Mie-particles). Assuming DHS for the dominant aC grains was found to provide the best general shape of the SED. All dust species are assumed to be in thermal contact.

The absorption and emission between 7\,\um\xspace and 10\,\um\xspace and around 14\,\um\xspace that is not fitted in our SED model can be explained by molecular bands of e.g. HCN and C$_2$H$_2$ \citep{gonzalez-alfonso1999,cernicharo1999_iso}.

\begin{figure}
\includegraphics[angle=0,width=\linewidth]{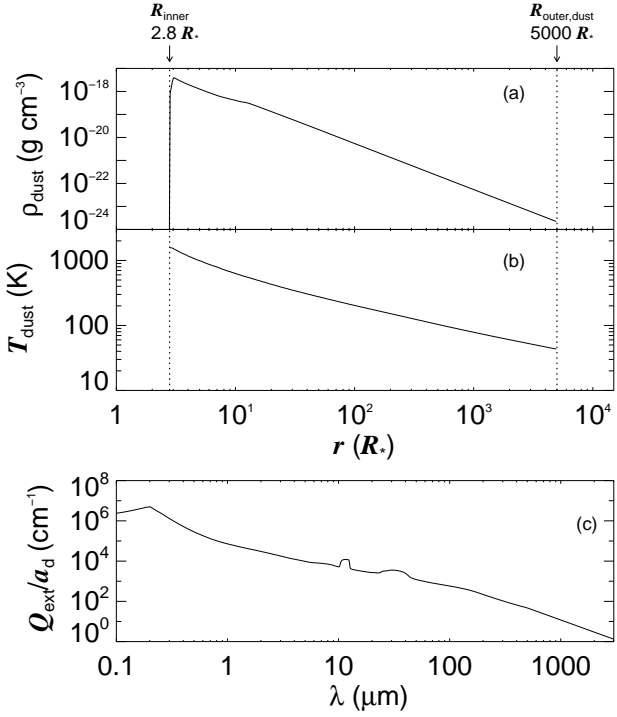}
\caption{Spatial distribution and properties of the dust for our best SED fit to the ISO data at $\varphi$=0.24 in Fig.~\ref{fig:sed}: \emph{(a)} dust density $\rho_\mathrm{dust}(r)$, \emph{(b)} dust temperature $T_\mathrm{dust}(r)$ , and \emph{(c)}  $Q_{\mathrm{ext}}/a_{\mathrm{d}}$. The dashed lines indicate the inner and outer radii of the dusty envelope.}\label{fig:dustprops}
\end{figure}

Based on an average of  the specific densities of the dust components $\rho_{\mathrm{s}}=2.41$\,g\,cm$^{-3}$, we find a dust mass-loss rate of $4.0\times10^{-8}$\,$M_{\sun}$\,yr$^{-1}$, with the dust density $\rho_\mathrm{dust}(r)$, dust temperature $T_\mathrm{dust}(r)$ and $Q_{\mathrm{ext}}/a_{\mathrm{d}}$, with $Q_{\mathrm{ext}}$ the total extinction efficiency, shown in Fig.~\ref{fig:dustprops}. The inner radius of the dusty envelope, i.e. the dust condensation radius of the first dust species to be formed, is determined at $R_\mathrm{inner}$=2.7\,$R_{\star}$ by taking into account pressure-dependent condensation temperatures for the different dust species \citep{kama2009}. These dust quantities are used as input for the gas radiative transfer, modelled in Sect.~\ref{sect:radtrangas}.

\begin{table}[ht]\caption{Dust composition in the CSE of \irc, used to produce the fit to the SED in Fig.~\ref{fig:sed}. The columns list the dust species, the assumed shapes, the respective mass fractions, and the references to the optical constants of the different dust species.}\label{tbl:dustcomposition}
\centering
\setlength{\tabcolsep}{1.5mm}
\begin{tabular}{lccl}
\hline\hline\\[-2ex]
Dust species		&Shape& Mass fraction 	&References\\
		&	&(\%)&\\
\hline\\[-2ex]
Amorphous carbon & DHS &53	&\cite{preibisch1993}\\
Silicon carbide	&CDE&25 &\cite{pitman2008}\\
Magnesium sulfide&CDE &22 &\cite{begemann1994}\\
\hline
\end{tabular}
\end{table}

\subsection{CO} \label{sect:radtrangas}
To constrain the gas kinetic temperature $T_\mathrm{kin}(r)$ and the gas density $\rho_\mathrm{gas}(r)$ throughout the envelope, we modelled the emission of 20 rotational transitions of $^{12}$CO (Sect.~\ref{sect:radtrangas}), measured with ground-based telescopes and \emph{Herschel}/HIFI. The gas radiative transfer is treated with the non-local thermal equilibrium (NLTE) code \texttt{GASTRoNOoM} \citep{decin2006_gastronoom,decin2010_gastronoom}. To ensure consistency between the gas and dust radiative transfer models, we combine \texttt{MCMax} and \texttt{GASTRoNOoM}, by passing on the dust properties (e.g. density and opacities) from the model presented in Sect.~\ref{sect:radtrandust} and Fig.~\ref{fig:dustprops} to the gas modelling. The general method behind this will be described in detail by \citet[\emph{in prep.}]{lombaert2011}.

\begin{figure} [ht]
\includegraphics[width=\linewidth]{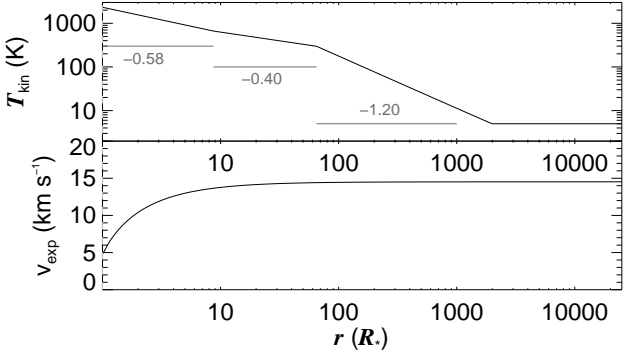}
\caption{Overview of the gas kinetic temperature and expansion velocity used in the radiative transfer model calculated with \texttt{GASTRoNOoM}.  In the top panel, we indicated the exponents $\alpha$ from the $T_{\mathrm{kin}}(r)\propto r^{\,\alpha}$-power laws used to describe the kinetic temperature, and the radial ranges they apply to.  See Sect.~\ref{sect:radtrangas}.}\label{fig:radtranprofile}
\end{figure}

The large data set of high-spectral-resolution rotational transitions of CO consists of 10 lines observed from the ground and 10 lines observed with HIFI, which are listed in Table~\ref{tbl:phaselstar}. Since the calibration of ground-based data is at times uncertain \citep{skinner1999}, large data sets of lines that are observed simultaneously and with the same telescope and/or instrument are of great value. Also, the observational uncertainties of the HIFI data are significantly lower than those of the data obtained with ground-based telescopes ($20-40\%$). Since the HIFI data of CO make up  half of the available high-resolution lines in our sample, these will serve as the starting point for the gas modelling.

The adopted CO laboratory data are based on the work of \cite{goorvitch1994} and \cite{winnewisser1997}, and are summarised in the Cologne Database for Molecular Spectroscopy \citep[CDMS;][]{mueller2005_cdms}. Rotational levels $J=0$ up to $J=60$ are taken into account for both the ground vibrational state and the first excited vibrational state. CO-H$_2$ collisional rates were adopted from \cite{larsson2002}.

To reproduce the CO lines within the observational uncertainties, we used $L_{\star}$=11300\,$L_{\sun}$, and a temperature profile\footnote{The central star is assumed to be a black body at a temperature $T_\mathrm{eff}$.} that is a combination of three power laws: $T_\mathrm{kin}(r)\propto r^{-0.58}$ for $r\leq9\,R_{\star}$, $T_\mathrm{kin}(r)\propto r^{-0.4}$ for $10\,R_{\star}\leq r\leq65\,R_{\star}$, and $T_\mathrm{kin}(r)\propto r^{-1.2}$ at larger radii, based on the work by \cite{fonfria2008} and \cite{decin2010_water}. The minimum temperature in the envelope is set to 5\,K. The radial profiles of $T_{\mathrm{kin}}(r)$ and $\varv(r)$ are shown in Fig.~\ref{fig:radtranprofile}. Using a fractional abundance CO/H$_2$ of $6\times10^{-4}$ in the inner wind, we find that a gas-mass-loss rate $\dot M$ of $1.5\times10^{-5}$\,$M_{\sun}$\,yr$^{-1}$ reproduces the CO lines very well. Combining the results from Sect.~\ref{sect:radtrandust} with those from the CO model, we find a gas-to-dust-mass ratio of 375, in the range of typical values for AGB stars \citep[$10^2-10^3$; e.g.][]{ramstedt2008}.

\begin{figure*} [ht]
\centering
\includegraphics[width=\linewidth]{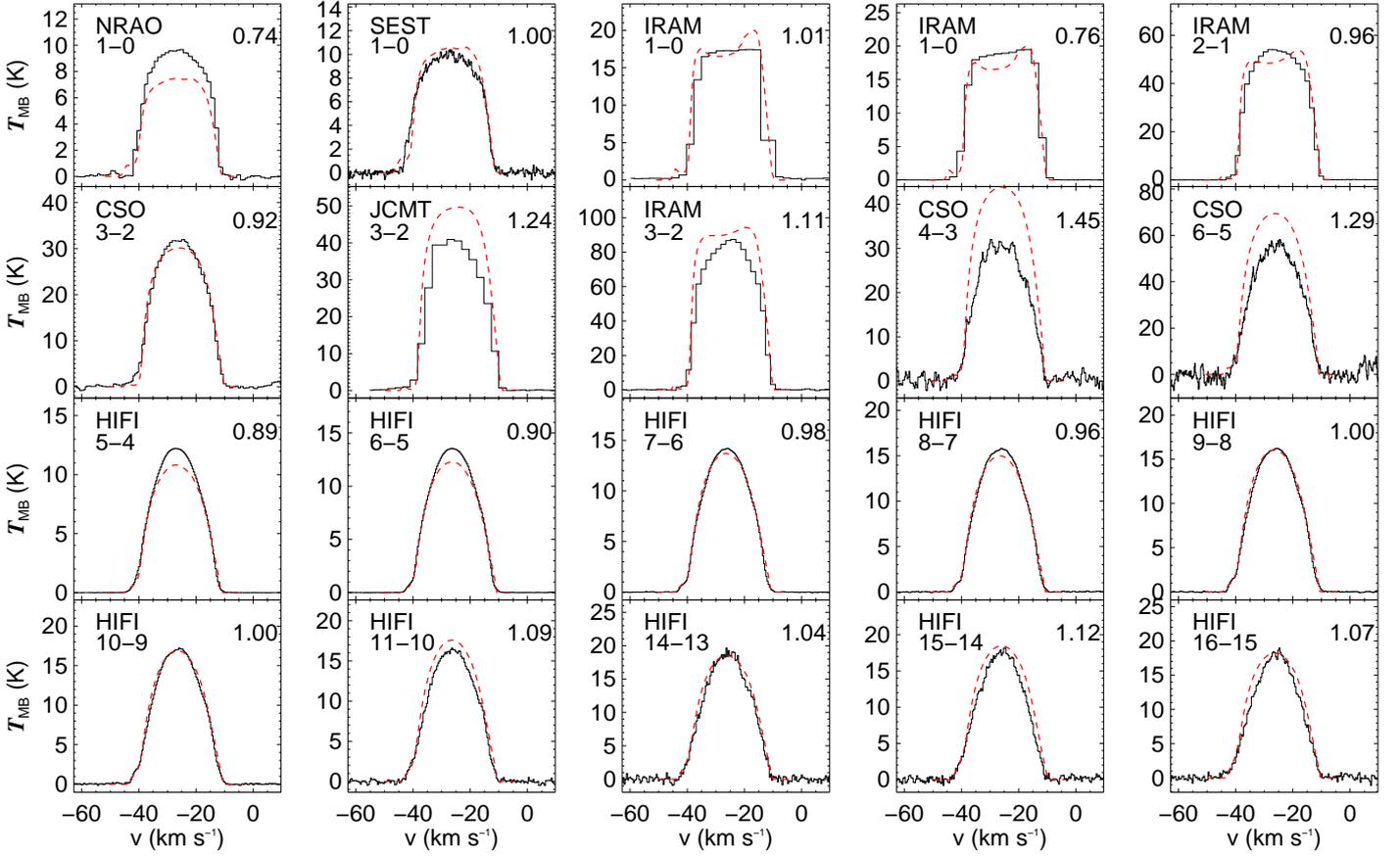}
\caption{Comparison of the observed $^{12}$CO emission lines \emph{(black histogram)}, and the line profiles predicted by the \texttt{GASTRoNOoM}-model \emph{(red dashed line)} with parameters as listed in Table~\ref{tbl:ircmodelparameters}. The line transitions $J-(J-1)$ and the telescopes with which they were observed are indicated in the upper left corner of every panel. The factor $I_\mathrm{MB,model}/I_\mathrm{MB,data}$ is given in the upper right corner of every panel.}\label{fig:colines}
\end{figure*}

The predicted CO line profiles are shown and compared to the observations in Fig.~\ref{fig:colines}. All lines are reproduced very well in terms of integrated intensity, considering the respective observational uncertainties. The fraction of the predicted and observed velocity-integrated main-beam intensities $I_\mathrm{MB,model}/I_\mathrm{MB,data}$ varies in the range $74-145\%$ for the ground-based data set, and in the range $89-109\%$ for the HIFI data set. The shapes of all observed lines are also well reproduced, with the exception of the IRAM lines (possibly due to excitation and/or resolution effects).

\begin{table} [t]
\centering
\caption{Overview of the transitions of CO and \ethynyl shown throughout this paper. We list the transitions, the telescope/instrument, the dates of observation, and the literature references in case of previously published data. These references are: (1) \cite{huggins1988_co}, (2) \cite{olofsson1993_co},  (3)  \citet[][\emph{in prep.}]{cernicharo2011_3mm}, (4) \cite{groenewegen1996_co}, (5) \cite{wang1994}, (6) \citet[][\emph{in prep.}]{kahane2011_1mm}, (7) \cite{teyssier2006}, (8) this paper, (9) \cite{decin2010_silicon},  (10) \cite{kahane1988}, (11) \cite{cernicharo2000_2mm}, (12) \cite{agundez2010_cnmin}, (13) \cite{cernicharo2011_hcnvib}. } \label{tbl:phaselstar}
\setlength{\tabcolsep}{2.45mm}
\begin{tabular} {cccc}
\hline\hline\\ [-2ex]
Transitions&Telescope&Date& Ref. \\[0.5ex]
\multicolumn{1}{l}{\textbf{$^{12}$CO($J\rightarrow J-1$)}}&\cline{1-3}\\
$1-0$&NRAO&Jun, 1986 &1\\
$1-0$&SEST &Oct 13, 1987&2\\
$1-0$&IRAM&Sept 3, 2004& 3\\
$1-0$&IRAM&Oct 8, 1991&4\\
$2-1$&IRAM&Aug 1, 2003&3\\
$3-2$&CSO&Jun, 1993&5\\
$3-2$&JCMT&Jul 13, 1992&4\\
$3-2$&IRAM&Feb 3, 2010&6\\
$4-3$&CSO& Feb 2002\tablefootmark{(a)} &7\\
$6-5$&CSO& Feb 2002\tablefootmark{(a)} &7\\
\quad$5-4$ \quad ... \quad $11-10$& HIFI&May 11-13, 2010&8\\
$14-13$ \quad ... \quad $16-15$&''&''&''\\
$14-13$ \quad ... \quad $25-24$&PACS& Nov 12, 2009&8, 9\\
$27-26$ \quad ... \quad $42-41$&''&''&''\\[2ex]
\multicolumn{1}{l}{\textbf{\ethynyl($N\rightarrow N-1$)}}&\cline{1-3}\\
$1-0$ & IRAM	& May 27, 1985 &10\\
$2-1$& IRAM		& Jan 7, 2002 &11\\
$3-2$&IRAM		&Jan 29, 2010 &12\\
$4-3$&IRAM		&Mar 28, 2010&6, 13\\
$6-5$&HIFI &May 11-13, 2010&8\\
$7-6$&''&''&''\\
$8-7$&''&''&''\\[0.5ex]
\hline
\end{tabular}
\tablefoot{\tablefoottext{a} \cite{teyssier2006} state that observations were carried out between September 2001 and February 2002.  A search of the CSO archive database resulted in observations in February 2002 that could be linked to their paper.}
\end{table}

\begin{figure*}[ht]
\centering
\includegraphics[angle=90,width=\linewidth]{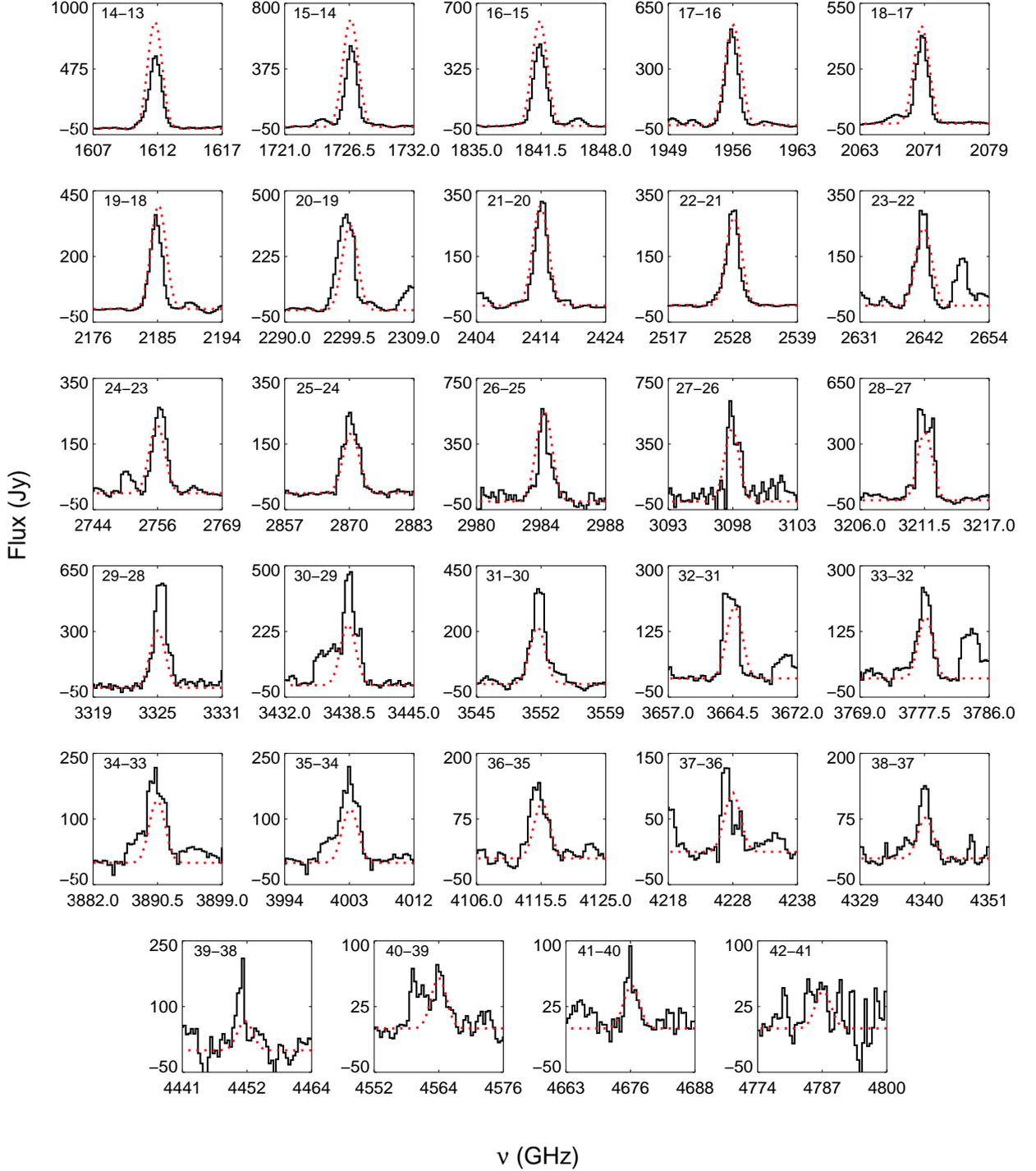}
\caption{Comparison of CO-emission lines measured with PACS \emph{(black histogram)} and the predicted line profiles \emph{(red dotted line)}. The transitions $J-(J-1)$ are labelled in the top left corner of each subpanel.}\label{fig:pacs}
\end{figure*}

\begin{table} [t]
\caption{Parameter set used to model the CO lines, as discussed in Sect.~\ref{sect:radtrangas}.}\label{tbl:ircmodelparameters}
\begin{center}
\setlength{\tabcolsep}{1.5mm}
\begin{tabular}{ll|ll}
\hline\hline\\[-2ex]
$d$ 	& 150\,pc			& $\dot M_\mathrm{dust}$\tablefootmark{(a)}	& $4.0\times10^{-8}$\,$M_{\sun}$\,yr$^{-1}$  \\
$L_{\star}$\tablefootmark{(a)}& 11300\,$L_{\sun}$&$\dot M_\mathrm{gas}$	& $1.5\times10^{-5}$\,$M_{\sun}$\,yr$^{-1}$\\
$R_{\star}$ 	& $4.55\times 10^{13}$\,cm &CO/H$_2$& $6\times 10^{-4}$\\ 
&	$\approx 655$\,$R_{\sun}$&$\varv_{\infty}$&14.5\,km\,s$^{-1}$	\\
$R_\mathrm{inner}$ \tablefootmark{(a)}	& 2.7\,$R_{\star}$&$\varv_{\mathrm{turb}}$&1.5\,km\,s$^{-1}$\\
$R_\mathrm{outer}$ & 25000\,$R_{\star}$ &$T_{\mathrm{kin}}(r)$ &$\propto r^{-0.58} \,(1\,R_{\mathrm{\star}} \leq r<\,\,9\,R_{\mathrm{\star}})$\\
$T_{\mathrm{eff}}$& 2330\,K	&&$\propto r^{-0.40} \,(9\,R_{\mathrm{\star}}\leq r<65\,R_{\mathrm{\star}})$ \\
&&&$\propto r^{-1.20} \,(65\,R_{\mathrm{\star}}\leq r)$ \\
&&&\\
\hline\\[-6ex]
\end{tabular}
\end{center}
\tablefoot{\tablefoottext{a}{From SED modelling.}}
\end{table}

Fig.~\ref{fig:pacs} shows the comparison between the observed CO transitions $J=14-13$ up to $J=42-41$ in the PACS spectrum and the modelled line profiles, convolved to the PACS resolution. %\textbf{Since the data show the total flux observed by all spaxels, no point-source correction \citep{poglitsch2010} was applied. A more detailed discussion on this topic is given in the (online) appendix.}
Considering the uncertainty on the absolute data calibration and the presence of line blends with e.g. HCN, C$_2$H$_2$, SiO, SiS, or H$_2$O in several of the shown spectral ranges \citep{decin2010_silicon}, the overall fit of the CO lines is very good, with exception of the lines $J=14-13,\,15-14$, and $16-15$. This significant difference between the model predictions and the data could, to date, not be explained. Both the HIFI data and the higher-$J$ PACS lines are reproduced very well by the model.
%, and it is unclear whether the discrepancy for these three lines is due to, e.g., a calibration issue\footnote{Cross-calibration efforts between the different \emph{Herschel} instruments are currently ongoing.}.  }

\begin{figure*}[ht]
\centering
\includegraphics[width=.75\linewidth]{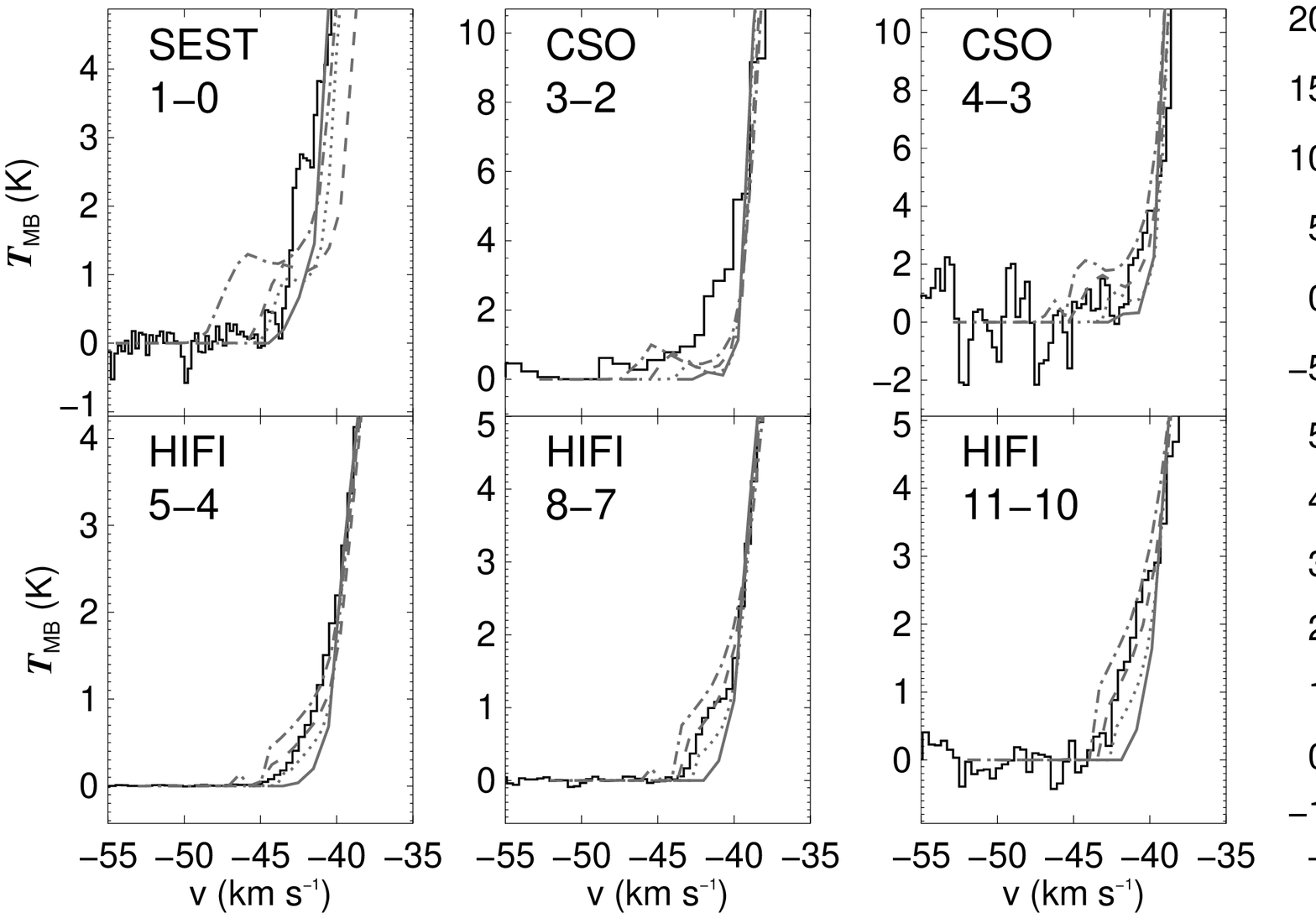}
\caption{Zoom on the blue wing of a selection of the CO line profiles, with the transitions $J-(J-1)$ labelled in the top left corner of each panel. We show the comparison between the observed lines and predicted lines with assumed values of $\varv_{\mathrm{turb}}$\,=\,0.5\,km\,s$^{-1}$ \emph{(full grey)}, 1.0\,km\,s$^{-1}$ \emph{(dotted grey)},\,the adopted value of 1.5\,km\,s$^{-1}$ \emph{(dashed grey)}, and 2.0\,km\,s$^{-1}$ \emph{(dash-dotted grey)}.}\label{fig:vturb}
\end{figure*}

\cite{skinner1999} noted the possible presence of a gradient in the turbulent velocity $\varv_{\mathrm{turb}}$, with decreasing values for increasing radial distance from the star. We find no clear evidence in our data set that this decrease in $\varv_{\mathrm{turb}}$ is present. The model shown in Fig.~\ref{fig:colines} uses $\varv_{\mathrm{turb}}$\,=\,1.5\,km\,s$^{-1}$, and reproduces the line shapes and intensities to a high degree. In Fig.~\ref{fig:vturb} we compare the predictions of a set of CO-lines using different values for $\varv_{\mathrm{turb}}$ in order to assess the influence of this parameter. \\
Assuming a lower value of 0.5\,km\,s$^{-1}$ or 1.0\,km\,s$^{-1}$ yields an incomplete reproduction of the  emission profile between $-44$\,km\,s$^{-1}$ and $-41$\,km\,s$^{-1}$ for all lines observed with HIFI. Assuming a higher value of 2\,km\,s$^{-1}$ leads to the overprediction of the emission in this $\varv$-range for all observed lines, with the strongest effect for the lowest-$J$ lines. The influence of the different values of the turbulent velocity on the emission in the $\varv$-range of $-41$ to $-12$\,km\,s$^{-1}$ is negligible.

%\clearpage

\section{\ethynyl}\label{sect:cch}
\subsection{Radical spectroscopy}\label{sect:spectroscopy}
\ethynyl ($\bullet\,\mathrm{C}\bond3\mathrm{C}\bond1\mathrm{H}$) is a linear molecule with an open shell configuration, i.e. it is a radical, with an electronic \twosigmaplus ground-state configuration \citep{mueller2000}. Spin-orbit coupling causes fine structure (FS), while electron-nucleus interaction results in hyperfine structure (HFS). Therefore, to fully describe \ethynyl in its vibrational ground-state, we need the following coupling scheme:
\begin{eqnarray}\label{eq:quantumrelations}
\qquad\vec{J}&=&\vec{N}+\vec{S}\\
\qquad\vec{F}&=&\vec{J}+\vec{I},
\end{eqnarray}
where $\vec{N}$ is the rotational angular momentum, not including the electron or quadrupolar angular momentum ($\vec{S}$ and $\vec{I}$, respectively), $\vec{J}$ is the total rotational angular momentum, and $\vec{F}$ is the nuclear spin angular momentum. 

The strong $\Delta J=\Delta N$ FS components have been detected up to $N=9-8$. Relative to the strong components, the weaker $\Delta J=0$ components decrease rapidly in intensity with increasing $N$, and they have been detected up to $N=4-3$. The HFS has also been (partially) resolved for transitions up to $N=4-3$.

Spectroscopic properties of the ground vibrational state of \ethynyl have most recently been determined by \cite{padovani2009} and are the basis for the entry in the CDMS.

Our treatment of the radiative transfer (Sect.~\ref{sect:cchradtran}) does not deal with the FS and HFS of \ethynyl, and is limited to the prediction of rotational lines $N \rightarrow N^{\prime}$ with $\Delta N=1$, according to a \onesigma approximation. Therefore, all levels treated are described with only one quantum number $N$. The final line profiles are calculated by splitting the total predicted intensity of the line over the different (hyper)fine components, depending on the relative strength of these components (CDMS), and preserving the total intensity. Note that, strictly speaking, this splitting is only valid under  LTE conditions, but that, due to our approach of \ethynyl as a $^{1}\Sigma$-molecule, we will apply this scheme throughout this paper.

\ethynyl has one bending mode \nutwo, and two stretching modes, $\nu_1$ and $\nu_3$. The rotational ground-state of the bending mode (\nutwo) is situated $\sim$530\,K above the ground state. The $\mathrm{C}\bond1\mathrm{C}$ stretch ($\nu_3$) at $\sim$2650\,K is strong. The $\mathrm{C}\bond1\mathrm{H}$ stretch ($\nu_1$) at $\sim$4700\,K is weak, and is resonant with the first excited $^2\Pi$ electronic state at $\sim$5750\,K of \ethynyl. For each of these vibrational states ---  ground-state, $\nu_2$, $\nu_3$, and $\nu_1$ --- we consider 20 rotational levels, i.e. ranging from $N=0$ up to $N=19$.

The equilibrium dipole moments of the ground and first excited electronic states have been calculated as $\mu=0.769$ and 3.004\,Debye, respectively \citep{woon1995}. The dipole moments of the fundamental vibrational states in the ground electronic state are also assumed to be 0.769\,Debye. This is usually a good assumption as shown in the case of HCN by \cite{deleon1984}. The band dipole moments for rovibrational transitions were calculated from infrared intensities published by \cite{tarroni2004}: $\mu$($\nu_2$=1$\rightarrow$0)\,=\,0.110\,Debye, $\mu$($\nu_3$=1$\rightarrow$0)\,=\,0.178\,Debye, and $\mu$($\nu_1$=1$\rightarrow$0)\,=\,0.050\,Debye. The influence of the vibrationally excited states on transitions in the vibrational ground state will be discussed in Sect.~\ref{sect:cchradtran}.

A more complete treatment of \ethynyl will likely have to take into account, e.g., overtones of the $\nu_2$-state. These have  non-negligible intrinsic strengths \citep{tarroni2004} because of anharmonicity and vibronic coupling with the first excited electronic state. Spectroscopic data for several of these are already available in the CDMS.  A multitude of high-lying states may also have to be considered as they have rather large intrinsic intensities because of the vibronic interaction between the ground and the first excited electronic states.

\begin{table}[ht]\centering\caption{Properties of the transitions $N\rightarrow N^{\prime}$ used to model the \ethynyl lines presented in Fig.~\ref{fig:cch_all}. $\nu$ is the rest frequency of the transition, $E_{N}/k$ is the energy level of the upper level $N$ of the transition divided by the Boltzmann constant $k$, $A_{N\rightarrow N^{\prime}}$ is the Einstein-$A$ coefficient of the transition, $S_{N \rightarrow N^{\prime}}$ is the theoretical line strength of the line. $g_N$ is the statistical weight of the upper level $N$. }\label{tab:cchmolparameters}
 \begin{tabular}{ccccccc}
\hline\hline\\[-2ex]
$N$	&$N^{\prime}$	&$\nu$	&$E_{N}/k$	&$A_{N\rightarrow N^{\prime}}$	&$S_{N \rightarrow N^{\prime}}$	&$g_N$	\\
	&		&(MHz)		&(K)		&(s$^{-1}$)				&				&				\\
\hline\\[-2ex]
1& 	0&   		 87348.635&      4.2&  		1.53$\times10^{-6}$ 			& 1 			&    3\\
2& 	1&   		174694.734&     12.6&  		1.47$\times10^{-5}$ 			& 2 			&    5\\
3& 	2&   		262035.760&     25.2&  		5.31$\times10^{-5}$ 			& 3 			&    7\\
4& 	3&   		349369.178&     41.9&  		1.30$\times10^{-4}$ 			& 4 			&    9\\
6& 	5&   		524003.049&     88.0&  		4.57$\times10^{-4}$ 			& 6 			&   13\\
7& 	6&   		611298.435&    117.4&  		7.34$\times10^{-4}$ 			& 7 			&   15\\
8& 	7&   		698576.079&    150.9&  		1.10$\times10^{-3}$ 			& 8 			&   17\\
\hline
\end{tabular}
\end{table}

\subsection{Observational diagnostics}\label{sect:cchobsdiagnostics}
The \ethynyl-emission doublets due to the molecule's fine structure (Sect.~\ref{sect:spectroscopy}) are clearly present in all high-resolution spectra shown in Fig~\ref{fig:cch_all}. For $N=1-0$, $2-1$, and $3-2$, the hyperfine structure is clearly detected in our observations. The double-peaked line profiles are typical for spatially resolved optically thin emission \citep[Olofsson, in][]{habing2003}, and indicate that the emitting material has reached the full expansion velocity\footnote{We assume that the half width at zero level of the CO emission lines, i.e. 14.5\,km\,s$^{-1}$, is indicative for the highest velocities reached by the gas particles in the CSE.}, i.e. $\varv_{\infty}=14.5$\,km\,s$^{-1}$. This is in accordance with the observed position of the emitting shells \citep[$\sim$16\arcsec; ][]{guelin1999} and the velocity profile derived for this envelope (Sect.~\ref{sect:radtrangas}). 

To derive information on the regime in which the lines are excited, we construct a rotational diagram \citep{schloerb1983}, using
\begin{equation}\label{eq:rotdiagram}
\qquad\frac{N}{Z(T_\mathrm{rot})}\times\exp\left(-\frac{E_u}{kT_\mathrm{rot}}\right) =\frac{3k\times10^{36}}{8\pi^3}\frac{I_{\mathrm{\varv,corr}}}{\nu S\mu^2},
\end{equation}
where $N$ is the column density, $Z$ the temperature dependent partition function, $E_u$ the energy of the upper level of the transition in cm$^{-1}$, $k$  the Boltzmann constant in units of cm$^{-1}$\,K$^{-1}$, $T_\mathrm{rot}$ the rotational temperature in K, $\mu$ the dipole moment in Debye, and $\nu$ the frequency of the transitions in Hz. $I_{\mathrm{\varv,corr}}$ is the velocity-integrated antenna temperature, 
\begin{equation}
\qquad I_{\mathrm{\varv,corr}}=\frac{1}{f_\mathrm{bff} \eta_\mathrm{MB}}\int_{\varv_\mathrm{LSR}-\varv_{\infty}}^{{\varv_\mathrm{LSR}+\varv_{\infty}}}{T_\mathrm{A}^{*} d\varv}
\end{equation}
corrected for the beam filling factor \citep{kramer1997}
\begin{equation}
f_\mathrm{bff}=
\left\lbrace\begin{array}{ll}
1-\exp\left(-\left[\sqrt{\ln2}\times\theta_\mathrm{source}/\theta_\mathrm{beam}\right]^2\right),&\theta_\mathrm{source}< \theta_\mathrm{beam}\\
1, &\theta_\mathrm{source}\geq \theta_\mathrm{beam}
\end{array}\right.
\end{equation}
and for the beam efficiency $\eta_\mathrm{MB}$ (see Table~\ref{tbl:iramhifi}). We have assumed a uniform emission source with a size of 32\arcsec\xspace in diameter for all \ethynyl transitions, based on the interferometric observations of the $N=1-0$ emission by \cite{guelin1999}.

\begin{figure}
\includegraphics[width=\linewidth]{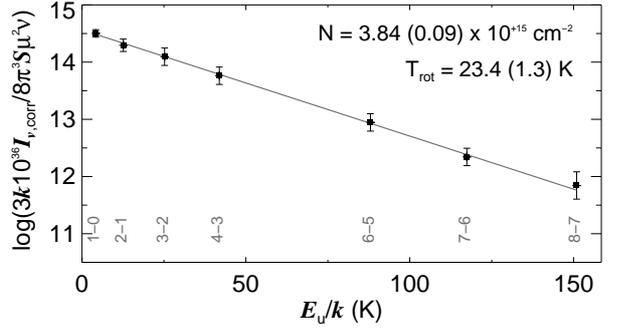}
\caption{Rotational diagram of the measured \ethynyl transitions, based on Eq.~\ref{eq:rotdiagram}. Transitions --- in the $^1\Sigma$ approximation --- are labelled in grey. The numbers given in parentheses are the uncertainties on the derived column density and rotational temperature.}\label{fig:rotdiagram}
\end{figure}

From Fig.~\ref{fig:rotdiagram} we see that the \ethynyl emission is characterised by a unique rotational temperature  $T_{\mathrm{rot}}=23.3(\pm1.3)$\,K, and a source-averaged column density $N=3.84(\pm0.09)\times 10^{15}$\,cm$^{-2}$. The unique rotational temperature points to excitation of the lines in one single regime, in accordance to the abundance peak in a confined area that can be characterised with one gas temperature \citep{guelin1999}. 

\subsection{Radiative transfer modelling}\label{sect:cchradtran}
\subsubsection{Collisional rates}
For rotational transitions within the vibrational ground state and within the $\nu_2=1$ bending state and the $\nu_1=1$ and $\nu_3=1$ stretching states we adopted the recently published collisional rates of HCN$-$H$_2$ by \cite{dumouchel2010}. The lack of  collisional rates for \ethynyl, and the similar molecular mass and size of HCN and \ethynyl inspire this assumption.  The collisional rates are given for 25 different temperatures between 5 and 500\,K, hence covering the range of temperatures relevant for the \ethynyl around \irc. Since $\nu_3$ and $\nu_1$ are at $\sim$2650\,K and $\sim$4700\,K, respectively, collisional pumping to these vibrationally excited levels is unlikely for the low temperatures prevalent in the excitation region of \ethynyl, i.e. the radical shell at $\sim$16\arcsec. Hence, for rovibrational transitions between the vibrationally excitated states ($\nu_2=1$, $\nu_3=1$, $\nu_1=1$) and the vibrational ground state  we adopted the same rates as for collisionally excited transitions within the vibrational ground state, but scaled down by a factor $10^{4}$, comparable to what has been done e.g. for H$_2$O by \cite{deguchi1990,maercker2008} and \cite{decin2010_gastronoom}. We did not consider collisionally excited transitions between the excited vibrational states. 

We note that recent calculations by \cite{dumouchel2010} have shown that the HNC collision rates at low temperatures differ by factors of a few from those of HCN. Since we may expect similar errors for \ethynyl, there is the need for accurate \ethynyl$-$H$_2$ collisional rates as well. The gas density, therefore, is not constrained better than this factor in the models presented in Sects.~\ref{sect:radtrangas} and~\ref{sect:spatialextent}.

\begin{figure}
\includegraphics[angle=0,width=\linewidth]{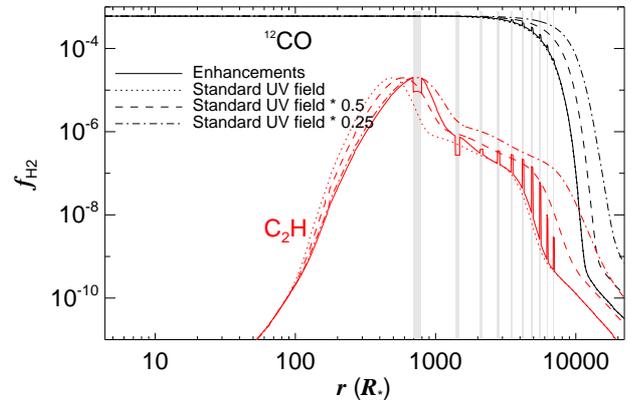}
\caption{Fractional abundance of $^{12}$CO (\emph{black}) and \ethynyl (\emph{red}) relative to H$_2$: model including the density enhancements \emph{(full lines)}, and models without density enhancements, assuming the average interstellar UV field of \cite{draine1978} scaled by a factor 1 \emph{(dotted lines)}, 0.5 \emph{(dashed lines)}, and 0.25 \emph{(dash-dotted lines)}. See Sect.~\ref{sect:spatialextent} for a discussion of this plot.}\label{fig:allabundances}
\end{figure}

\subsubsection{Constraining the \ethynyl abundance profile} \label{sect:spatialextent}
The calculation of the \ethynyl/H$_2$ fractional abundance is based on the chemical model discussed by \cite{agundez2010_cnmin}, assuming the average interstellar UV field from \cite{draine1978}, and a smooth envelope structure. The CO and \ethynyl abundances corresponding to the envelope model presented in Sect.~\ref{sect:radtrangas} are shown in Fig.~\ref{fig:allabundances}.

\begin{figure*}[ht]
\centering
\subfigure[Chemical model assuming time-independent mass loss\label{fig:cchmap_smooth_draine1}]
{\includegraphics[width=.45\linewidth]{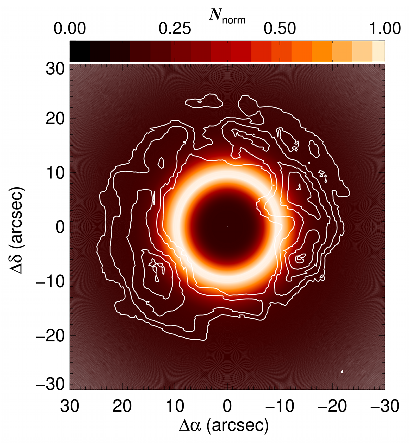}}
\subfigure[Chemical model assuming time-dependent mass loss\label{fig:cchmap_enhanced}]
{\includegraphics[width=.45\linewidth]{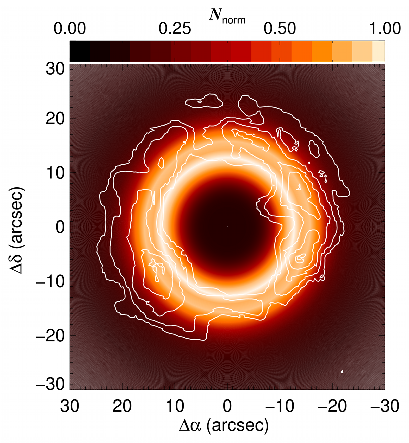}}
\subfigure[Chemical model assuming time-independent mass loss and an interstellar UV field a quarter as strong as in the standard model presented in panel (a)\label{fig:cchmap_smooth_draine4}]
{\includegraphics[width=.45\linewidth]{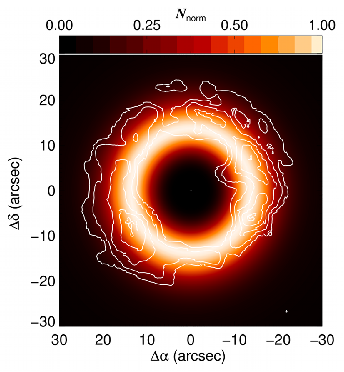}}
\caption{Contours (\emph{in white}) of the brightness distribution of the \ethynyl $N=1-0$ transition measured by \cite{guelin1999} with PdBI, overlayed on the normalised predicted brightness distribution, assuming LTE conditions. The three panels represent the results obtained with three different chemical models. For the map in panel (a), we assumed a constant mass loss, and the interstellar UV field of \cite{draine1978}. The map in panel (b) shows the brightness distribution when assuming density enhancements in the CSE as described by \cite{cordiner2009} and as discussed in Sect.~\ref{sect:spatialextent}. For panel (c), we assumed again a constant mass loss, but an interstellar UV field that is only 25\% as strong as that presented by \cite{draine1978}. \label{fig:cchmap}}
\end{figure*}

The Plateau de Bure Interferometer (PdBI) maps of \cite{guelin1999} show that the 3mm-lines of the three radicals \ethynyl, C$_4$H, and C$_6$H have their brightness peaks at a radial angular distance of $\sim$16\arcsec\xspace from the central star. In Fig.~\ref{fig:cchmap_smooth_draine1} the contours of the \ethynyl map by \cite{guelin1999} are overlayed on the normalised brightness distribution of the $N=1-0$ transition predicted by a (one-dimensional) LTE model based on the abundance profile mentioned above. The extracted PdBI contours correspond to the velocity channel at the systemic velocity of \irc, representing the brightness distribution of the \ethynyl-transition in the plane of the sky. From Fig.~\ref{fig:cchmap_smooth_draine1} it is clear that this model leads to a brightness distribution that is concentrated too far inwards to be in agreement with the shown PdBI contours. 

Several authors have established that the mass-loss process of \irc shows a complex time-dependent behaviour. The images of \cite{mauron1999} and \cite{leao2006} in dust-scattered stellar and ambient (optical) light, and the CO($J=1-0$) maps of \cite{fong2003} show enhancements in the dust and the gas density, with quasi-periodic behaviour. Recently, \cite{decin2011_shells} showed the presence of enhancements out to $\sim$320\arcsec\xspace based on PACS-photometry, consistent with the images of \cite{mauron1999}. 

\cite{brown2003} presented a chemical envelope model incorporating the dust density enhancements observed by \cite{mauron1999}. \cite{cordiner2009} added density enhancements in the gas at the positions of the dust density enhancements. They assumed complete dust-gas coupling, based on the work by \cite{dinh-v-trung2008} who compared maps of the molecular shells of HC$_3$N and HC$_5$N with the images by \cite{mauron1999}. To mimic these density enhancements in our model and evaluate the impact on the excitation and emission distribution of \ethynyl, we use the approximation of \cite{cordiner2009}, adding (to our ``basic'' model, presented in Sect.~\ref{sect:radtrangas}) ten shells of 2\arcsec\xspace width, with an intershell spacing of 12\arcsec, starting at 14\arcsec\xspace from the central star. These shells --- located at 14\arcsec--16\arcsec, 28\arcsec--30\arcsec, etc. --- are assumed to have been formed at gas-mass-loss rates of a factor six times the rate in the regions of ``normal density'', the intershell regions. We note that similar episodic mass loss was not added to the ``basic'' dust model presented in Sect~\ref{sect:radtrandust}. The uncertainties on the data presented in Sect.~\ref{sect:radtrandust} and the photometric variability of \irc are too large to constrain the expected small effect of the inclusion of these enhancements.

These density enhancements are included in the chemical model of \cite{agundez2010_cnmin} by combining two regimes. Firstly, a model is run in which the chemical composition of a parcel of gas is followed as it expands in the envelope, considering an augmented extinction ($A_V$) contribution from the density-enhanced shells located in the outer CSE. The density of the parcel is determined by the non-enhanced mass-loss rate ($\dot{M}$). A second model is run in which the composition of a density-enhanced shell is followed as it expands. The final abundance profile follows the abundance from the first model for the non-enhanced regions and follows the abundance of the second model for the density-enhanced shells. The resulting abundance profiles for $^{12}$CO and \ethynyl are compared to those corresponding to a smooth model without enhancements in Fig.~\ref{fig:allabundances}. The  $^{12}$CO-abundance is affected by the inclusion of density enhancements only in the outer regions of the CSE. The impact on the radiative-transfer results for $^{12}$CO is negligible, with differences in modelled integrated intensities of at most 3\%  compared to the model from Sect.~\ref{sect:radtrangas}. 

We find that the presence of density enhancements significantly alters the abundance profile of \ethynyl, as was also discussed by \cite{cordiner2009}. The abundance peak at $\sim$500\,$R_{\star}$ in the smooth model, is shifted to $\sim$800\,$R_{\star}$ in the enhanced model, as shown in Fig.~\ref{fig:allabundances}. This shift indicates that the photochemistry, in this particular case the photodissociation of \acetylene, is initiated at larger radii than in the smooth model. This is caused by the stronger extinction due to the density enhancements in the outer envelope. At the position of the two innermost density enhancements the fractional \ethynyl abundance is lower than in the intershell regions. The absolute abundance of \ethynyl in these regions, however, is higher by a factor $2-3$ than in the neighbouring intershell regions. This is a combined effect of \emph{(1)} the augmented shielding of C$_2$H$_2$ from incident interstellar UV radiation (which is no longer true for the outer shells), and \emph{(2)} chemical effects such as faster reactions with e.g. C$_2$H$_2$ to form larger polyynes.

We note that the correspondence between the interferometric observations and the modelled number densities has significantly improved by including density enhancements (Fig.~\ref{fig:cchmap_enhanced}). However, the peak intensity is located at somewhat too small radii compared to the observations. The adopted enhancements in the envelope are linked to $\dot{M}$-values of a factor six times as high as for the intershell regions, in correspondence with the values used in the model of \cite{cordiner2009}. Increasing this factor or the number of shells in our model would increase the total mass of material shielding the C$_2$H$_2$ molecules from photodissociation by photons from the ambient UV field, and would hence shift the peak intensity of the predicted emission outwards. Furthermore, the presence of numerous dust arcs out to $\sim$320\arcsec\xspace is reported by \cite{decin2011_shells} and their effect will be included in future chemical models.

The ``basic'' model from Sects.~\ref{sect:radtrangas} and \ref{sect:cchradtran} and the above introduced model with the density enhancements assume the interstellar UV field of \cite{draine1978}. To assess the effect of a weaker UV field on the photodissociation of C$_2$H$_2$, and hence on the spatial extent of \ethynyl, we tested models assuming interstellar UV fields weaker by a factor of 2 and 4. The corresponding \ethynyl abundance profiles are shown in Fig.~\ref{fig:allabundances}, and the the modelled $N=1-0$ brightness distribution for the latter case is shown in Fig.~\ref{fig:cchmap_smooth_draine4}. Although the correspondence in this figure is very good, the C$_4$H and C$_6$H abundances produced by this chemical model cannot account for the coinciding brightness peaks of \ethynyl, C$_4$H, and C$_6$H as reported by \cite{guelin1999}. In contrast, the density-enhanced model does reproduce this cospatial effect.

\subsubsection{Excitation analysis}
\begin{figure}[h]
\includegraphics[width=\linewidth]{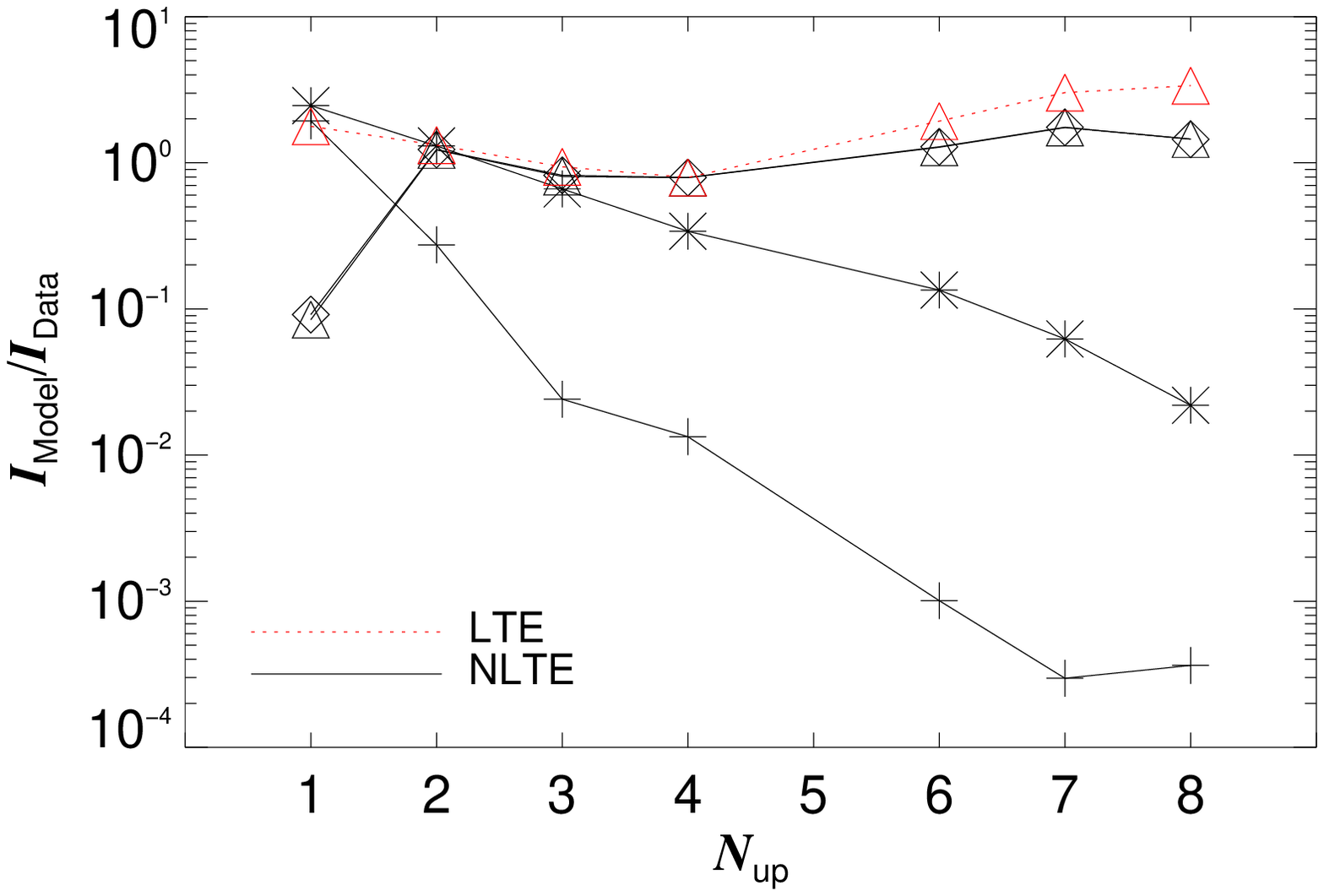}
\caption{Overview of the ratio $I_{\mathrm{Model}}/I_{\mathrm{Data}}$ of integrated intensities for the \ethynyl lines in Fig.~\ref{fig:cch_all}. The x-axis is labelled according to the upper $N$-level of the transition, $N_{\mathrm{up}}$. Different $^1\Sigma$-approximations of the \ethynyl molecule are plotted with different symbols: crosses ($+$) for the model including only the ground state (GS), asterisks ($*$) for the model including GS and the $\nu_2$ state, diamonds ($\diamond$) for the model including GS, $\nu_2$, and $\nu_3$, and triangles ($\triangle$) for the model including the GS and all three vibrationally excited states $\nu_2$, $\nu_3$, and $\nu_1$. Black symbols connected with full lines, and red symbols connected with dotted lines represent NLTE and LTE models, respectively.}\label{fig:c2happrox}
\end{figure}

\begin{figure*}[ht]
\centering
\subfigure[Lines observed with IRAM]{\includegraphics[width=.4\linewidth]{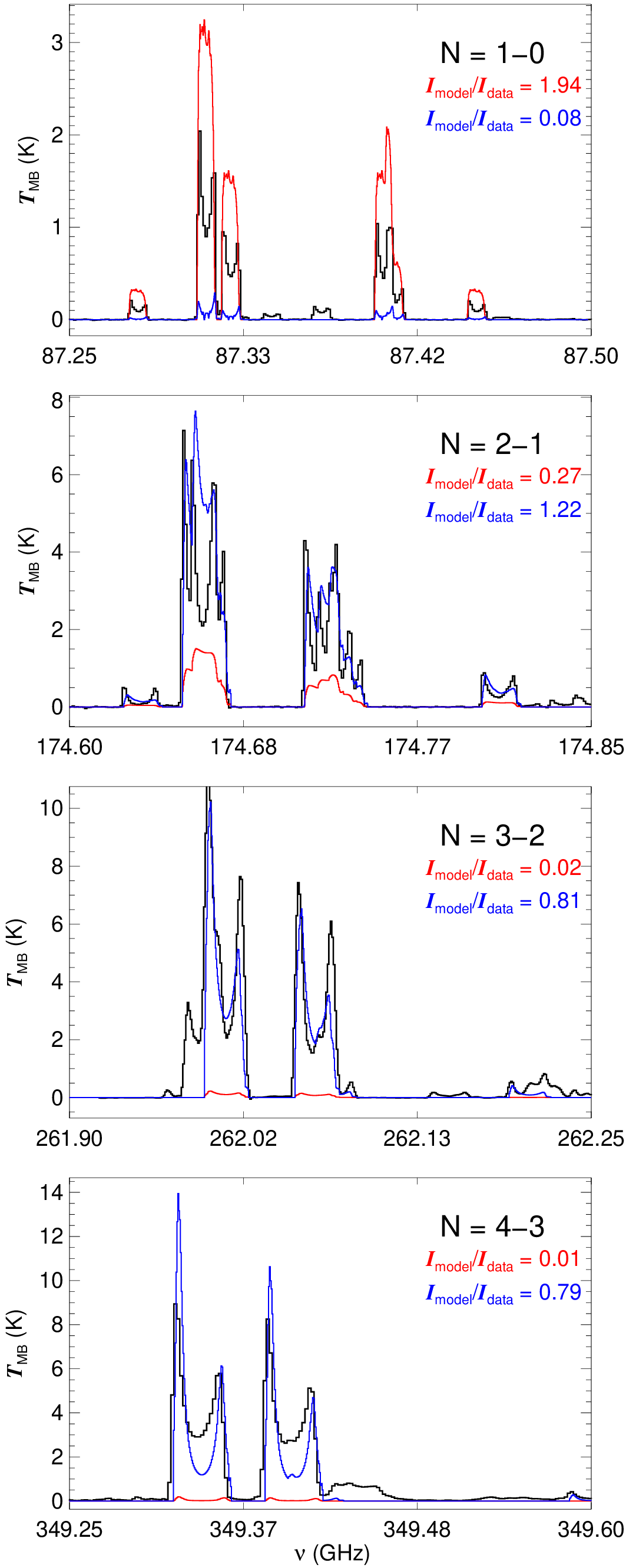}}
\hspace{.1\linewidth}
\subfigure[Lines observed with HIFI]{\includegraphics[angle=90,width=.4\linewidth]{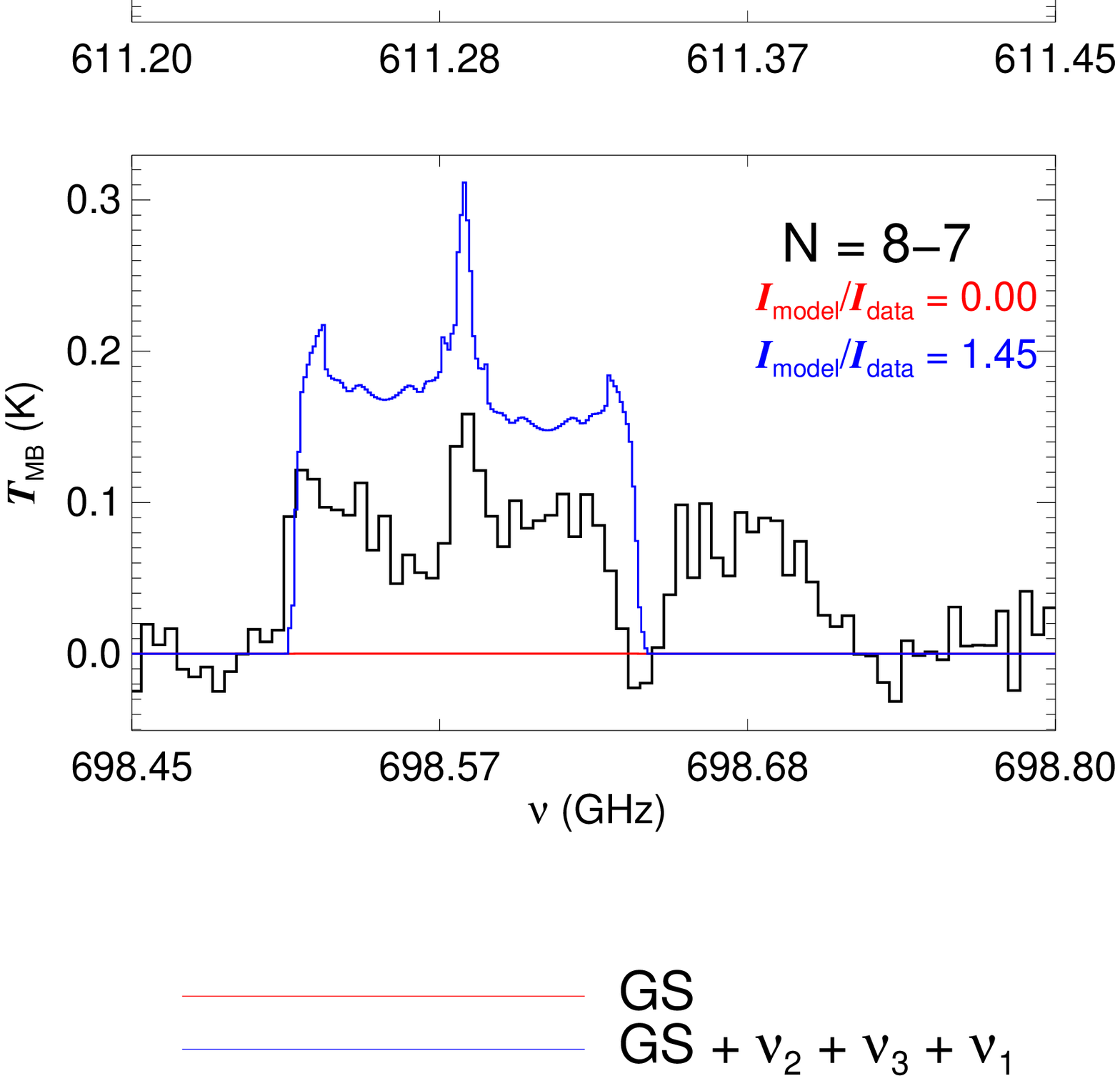}}
\caption{Comparison of the measured \ethynyl spectra \emph{(black histograms; see Sect.~\ref{sect:observations} and Fig.~\ref{fig:cch_all} for a description and for the identification of additional spectral features}) and GASTRoNOoM model predictions under NLTE conditions for the case where \emph{(red)} only the ground state is included, and \emph{(blue)} the ground state, and the three vibrational modes are taken into account. These predictions are based on the ``enhanced'' abundance profile shown in Fig.~\ref{fig:allabundances} and the $^1\Sigma$-approximation of \ethynyl. The transition $N-(N-1)$ and the ratio $I_{\mathrm{model}}/I_{\mathrm{data}}$ are stated in the upper right corner of each panel, following to the colour code of the plots. }\label{fig:cchpredictions}
\end{figure*}

Assuming the \ethynyl abundance obtained from the envelope model with density enhancements, we tested the influence of including the vibrational modes by modelling four cases: \emph{(1)} including only the ground-state (GS) level, \emph{(2)} including GS and $\nu_2$ levels, \emph{(3)} including GS, $\nu_2$, and $\nu_3$ levels, and \emph{(4)} including GS, $\nu_2$, $\nu_3$, and $\nu_1$ levels.  An overview of the ratio of the predicted and the observed integrated intensities, $I_{\mathrm{Model}}/I_{\mathrm{Data}}$, for these four cases is shown in Fig.~\ref{fig:c2happrox}. The comparison of the line predictions for case \emph{(1)} and case \emph{(4)} under NLTE conditions is shown in Fig.~\ref{fig:cchpredictions}.

\begin{figure}
\centering
\includegraphics[width=\linewidth]{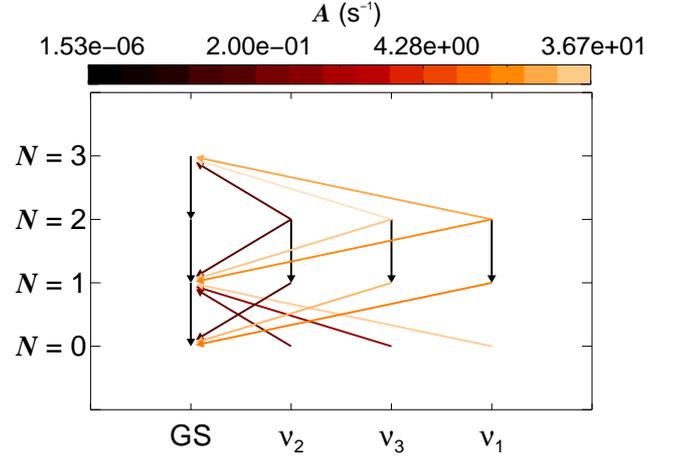}
\caption{Visualisation of the transitions responsible for the depopulation of the $N=1$ level in the vibrational ground state (GS), where vibrational states are indicated on the horizontal axis, and  rotational levels $N$ on the vertical axis. The colour of the arrows representing the transitions is indicative of the magnitude of the Einstein-$A$ coefficients of these transitions, as given by the colour bar.}\label{fig:ccheinstein}
\end{figure}

Under LTE conditions, $I_{\mathrm{Model}}/I_{\mathrm{Data}}$ for a given transition is the same for all four cases, indicating that the vibrationally excited states are not populated under the prevailing gas kinetic temperatures of $\sim$20\,K in the radical shell. Under NLTE conditions, however, the predicted line intensities are very sensitive to the inclusion of the vibrationally excited states. In the two cases where the $\nu_3$-state is included, the predicted intensity of the $N=1-0$ transition is only 9\% of the observed value, while transitions with $N_{\mathrm{up}}\geq4$ are more easily excited. This is linked to the involved Einstein-$A$ coefficients. For example, Table~\ref{tbl:einsteinacoeff} gives the Einstein-$A$ coefficients of the transitions involving the $N=2$ level of the vibrationally excited states and the $N=3$ and $N=1$ levels of the ground state. The higher values of the Einstein-$A$ coefficients for transitions to the $N=3$ level lead to a more effective population of this level than of the $N=1$ level. In particular, when including $\nu_3$ and $\nu_1$, this causes an underprediction of the intensity of the $N=1-0$ ground state transition, while this effect is insignificant when  only $\nu_2$ is included. A visualisation of this scheme is shown in Fig.~\ref{fig:ccheinstein}. This pumping mechanism is not limited to $N=3$, but also affects higher levels, explaining the easier excitation of the transitions with $N_{\mathrm{up}}\geq4$ mentioned before. 

\begin{table}[h]\caption{Einstein-$A$ coefficients (in s$^{-1}$) for the transitions involving the $N=1$ and $N=3$ levels of the ground state (GS).}\label{tbl:einsteinacoeff}
\centering
\begin{tabular}{ccll}
\hline\hline
\multicolumn{2}{c}{} &\multicolumn{2}{c}{lower level}\\
																												&&GS, $N=3$    & GS, $N=1$\\\hline\multicolumn{4}{c}{}\\[-2ex]
\multicolumn{1}{c}{}					&$\nu_2,\,N=2$   	                           &$1.09\times10^{-1}$		 	  &$8.16\times10^{-2}$\\
\multicolumn{1}{c}{upper level}&$\nu_3,\,N=2$		                             &$3.67\times10^{1}$			 &$2.50\times10^{1}$\\
\multicolumn{1}{c}{}					&$\nu_1,\,N=2$		                           &$1.68\times10^{1}$			   &$1.13\times10^{1}$\\\hline
\end{tabular}
\end{table}

\subsubsection{Turbulent velocity} 
As was done for CO, we tested the influence of $\varv_{\mathrm{turb}}$ on the predicted \ethynyl-emission, considering values $\varv_{\mathrm{turb}}$=0.5, 1.0, 1.5, and 2.0\,km\,s$^{-1}$.  We find that the predicted line intensity increases\footnote{This is valid for all lines, except for the $N=1-0$ transition, where the predicted intensity decreases by 40\%.} with $25-55\%$ when $\varv_{\mathrm{turb}}$ is increased from 0.5\,km\,s$^{-1}$ to 2.0\,km\,s$^{-1}$. From the comparison of the predicted and observed shapes of the emission lines, we conclude that values of $\varv_{\mathrm{turb}}$ in the range $0.5-1.5$\,km\,s$^{-1}$ give the best results. This is consistent with our discussion in Sect.~\ref{sect:radtrangas} for the CO-lines, and with values generally used for AGB envelopes \citep{skinner1999}.

\subsection{Vibrationally excited states}\label{sect:cchvib}
The removal of a hydrogen atom from C$_2$H$_2$ through photodissociation causes the produced \ethynyl molecule to be bent, rather than linear \citep{mordaunt1998}. This means that this formation route of \ethynyl favours the population of the \nutwo$=1$ state over the population of the ground state. However, the Einstein-$A$ coefficients for rovibrational transitions in the band $\nu_2=1\rightarrow0$ are in the range $0.01-1$\,s$^{-1}$, and by far exceed the photodissociation rates of C$_2$H$_2$, which are of the order of $10^{-9}$\,s$^{-1}$ \citep{vandishoeck2006}. Hence, we did not take this $\nu_2$-state population effect into account.

\begin{figure}
\centering
\includegraphics[width=.75\linewidth]{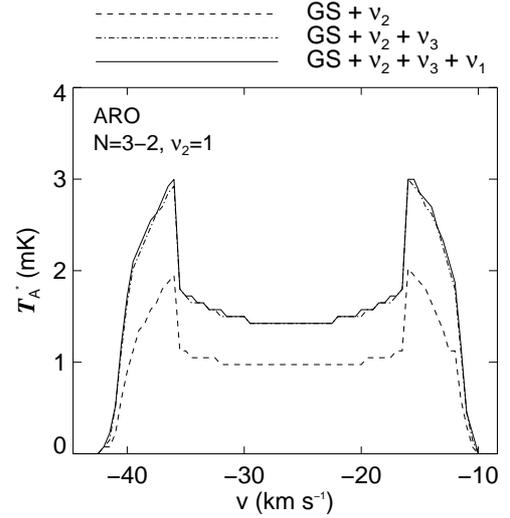}
\caption{The predicted line profile for the $N=3-2$ transition in the $\nu_2=1$ vibrationally excited mode, under assumption of NLTE conditions, and different $^{1}\Sigma$-approximations, as it would be observed with the Arizona Radio Observatory Submillimeter Telescope.}\label{fig:cchvibline}
\end{figure}

Recently, \cite{tenenbaum2010} reported on the detection of the \ethynyl $N=3-2$ transition in the \nutwo$=1$ state, with the Arizona Radio Observatory Submillimeter Telescope (ARO-SMT). Observed peak intensities are of the order of $\sim$$5-10$\,mK in antenna temperature, with reported noise levels around 3\,mK, and a total velocity-integrated intensity of $\sim$0.2\,K\,km\,s$^{-1}$. The $\nu_2=1,\,N=3$ level is located at $\sim$560\,K and is not populated under LTE conditions, given the gas-kinetic temperature in the radical shell ($\sim$20\,K). Under NLTE conditions, however, the $\nu_2$-levels are easily populated. The prediction of the $\nu_2=1,\,N=3-2$ emission under the different $^1\Sigma$-approximations is shown in Fig.~\ref{fig:cchvibline}. Under the $^1\Sigma$-approximation, and assuming an HPBW of 29\arcsec\xspace at 261\,GHz for ARO, we predict a U-shaped profile with peak antenna temperature $\sim$3\,mK. Taking into account the observed fine structure, we hence predict an integrated intensity which is a factor $\sim$4 lower than the observed value. Considering the substantial uncertainties (the low signal-to-noise ratio of the ARO observations, and the limited \ethynyl approximation in our model), this agreement is satisfactory.


\section{Summary}\label{sect:conclusions}
We presented new data of CO and \ethynyl obtained with HIFI, PACS, SPIRE and the IRAM 30\,m telescope. High-resolution spectra of CO transitions up to $J=16-15$, and of \ethynyl transitions up to $N=9-8$ were presented and modelled. The HIFI data of both CO and \ethynyl are the first high-frequency-resolution detections of these lines. They were obtained in the framework of a spectral survey of \irc over the complete frequency range of the HIFI instrument \citep{cernicharo2010_hifi1b,cernicharo2011_scan}.

From an SED fit to ISO data, PACS data, and a set of photometric points, covering the wavelength range $0.1-1000$\,\um, we obtained a dust-mass-loss rate of $4.0\times10^{-8}\,M_{\sun}$\,yr$^{-1}$ and a luminosity of 11300\,$L_{\sun}$ at a distance of 150\,pc. This luminosity value is valid at $\varphi=0.24$, the phase at which the ISO data were obtained. The luminosity is then expected to vary between 6250\,$L_{\sun}$ and 15800\,$L_{\sun}$ throughout the star's pulsational cycle, which has a period $\sim$649 days. 

In order to model \irc's wind, we performed the radiative transfer of the dusty component of the CSE consistently with the gas-radiative transfer of CO. The set of 20 high-spectral-resolution CO lines was modelled in order to constrain the physical parameters of \irc's CSE. The kinetic temperature throughout the envelope was described by previous results reported by \cite{fonfria2008} and \cite{decin2010_water}, and is now further constrained by the combination of all the high-resolution CO lines. The temperature profile is characterised by $T_\mathrm{kin}(r)\propto r^{-0.58}$ for $r\leq9\,R_{\star}$, $T_\mathrm{kin}(r)\propto r^{-0.4}$ for $10\,R_{\star}\leq r\leq65\,R_{\star}$, and $T_\mathrm{kin}(r)\propto r^{-1.2}$ at larger radii, with an effective temperature $T_\mathrm{eff}$=2330\,K. The derived mass-loss rate is $1.5\times10^{-5}$\,$M_{\sun}$\,yr$^{-1}$. This is consistent with earlier results \citep[e.g.][]{cernicharo2000_2mm} and gives a gas-to-dust-mass ratio of 375, in line with typical values stated for AGB stars \citep[e.g.][]{ramstedt2008}. Furthermore, we showed a very good agreement between the predictions for CO lines up to $J=42-41$ and the newly calibrated PACS spectrum of \irc. It is the first time that such a large coverage of rotational transitions of CO is modelled with this level of detail.

We extended our envelope model with the inclusion of episodic mass loss, based on the model of \cite{cordiner2009}. This assumption proved very useful in reconciling the modelled \ethynyl emission with the PdBI map of the $N=1-0$ transition of \ethynyl published by \cite{guelin1999}, and in reproducing the observed line intensities. A decrease of a factor four in the strength of the interstellar UV field also leads to a satisfactory reproduction the PdBI map, but resulted in  poorly modelled line intensities. The inclusion of density enhancements in \irc's CSE is further supported by observational results based on maps of dust-scattered light \citep{mauron1999}, molecular emisson \citep{fong2003}, and photometric maps recently obtained with PACS \citep{decin2011_shells}.

The ground-based observations of \ethynyl transitions involve rotational levels up to $N=4$ with energies up to $\sim$17\,cm$^{-1}$, corresponding to temperatures $\sim$25\,K. This temperature is close to the gas kinetic temperature at the position of the radical shell ($\sim$20\,K). The recent detection of strong \ethynyl-emission involving levels with energies up to $\sim$150\,K, however, calls for an efficient pumping mechanism to these higher levels. Due to the spectroscopic complexity of \ethynyl, with the presence of fine structure and hyperfine structure, we approximated the molecule as a $^1\Sigma$-molecule, exhibiting pure rotational lines, without splitting. We illustrated that the inclusion of the bending and stretching modes of \ethynyl is crucial in the model calculations, since high-energy levels are much more efficiently (radiatively) populated in this case. At this point, we have not yet  included overtones of the vibrational states, nor did we treat the resonance between vibrational levels in the electronic ground state and the first electronically excited $A^2\Pi$-state. Applying our simplified molecular treatment of \ethynyl, we can explain the strong intensities of the rotational lines in the vibrational ground state, except for the $N=1-0$ transition. We are also able to account for the excitation of the recently observed rotational transition in the $\nu_2=1$ state, showing the strength of our approach.

\begin{acknowledgements}
The authors wish to thank B. L. de Vries for calculating and providing dust opacities based on optical constants from the literature.

EDB acknowledges financial support from the Fund for Scientific Research - Flanders (FWO) under grant number G.0470.07. M.A is supported by a Marie Curie Intra-European Individual Fellowship within the European Community 7th Framework under grant agreement No. 235753. LD acknowledges financial support from the FWO.  JC thanks the Spanish MICINN for funding support under grants AYA2006-14876, AYA2009-07304 and CSD2009-03004. HSPM is very grateful to the Bundesministerium f\"ur Bildung und Forschung (BMBF) for financial support aimed at maintaining the Cologne Database for Molecular Spectroscopy, CDMS. This support has been administered by the Deutsches Zentrum f\"ur Luft- und Raumfahrt (DLR). MG and PR acknowledge support from the Belgian Federal Science Policy Office via de PRODEX Programme of ESA. RSz and MSch ackowledge support from grant N203 581040 of National
Science Center.

The Herschel spacecraft was designed, built, tested, and launched under a contract to ESA managed by the Herschel/Planck Project team by an industrial consortium under the overall responsibility of the prime contractor Thales Alenia Space (Cannes), and including Astrium (Friedrichshafen) responsible for the payload module and for system testing at spacecraft level, Thales Alenia Space (Turin) responsible for the service module, and Astrium (Toulouse) responsible for the telescope, with in excess of a hundred subcontractors.

HIFI has been designed and built by a consortium of institutes and university departments from across Europe, Canada and the United States under the leadership of SRON Netherlands Institute for Space Research, Groningen, The Netherlands and with major contribution¡s from Germany, France and the US. Consortium members are: Canada: CSA, U.Waterloo; France: CESR, LAB, LERMA, IRAM; Germany: KOSMA, MPIfR, MPS; Ireland, NUI Maynooth; Italy: ASI, IFSI-INAF, Osservatorio Astrofisico di Arcetri-INAF; Netherlands: SRON, TUD; Poland: CAMK, CBK; Spain: Observatorio Astronómico Nacional (IGN), Centro de Astrobiolog\'ia (CSIC-INTA). Sweden: Chalmers University of Technology - MC2, RSS \&  GARD; Onsala Space Observatory; Swedish National Space Board, Stockholm University - Stockholm Observatory; Switzerland: ETH Zurich, FHNW; USA: Caltech, JPL, NHSC.

SPIRE has been developed by a consortium of institutes led by Cardiff University (UK) and including Univ. Lethbridge (Canada); NAOC (China); CEA, LAM (France); IFSI, Univ. Padua (Italy); IAC (Spain); Stockholm Observatory (Sweden); Imperial College London, RAL, UCL-MSSL, UKATC, Univ. Sussex (UK); and Caltech, JPL, NHSC, Univ. Colorado (USA). This development has been supported by national funding agencies: CSA (Canada); NAOC (China); CEA, CNES, CNRS (France); ASI (Italy); MCINN (Spain); SNSB (Sweden); STFC (UK); and NASA (USA).

PACS has been designed and built by a consortium of institutes and university departments from across Europe under the leadership of Principal Investigator Albrecht Poglitsch located at Max-Planck-Institute for Extraterrestrial Physics, Garching, Germany. Consortium members are: Austria: UVIE; Belgium: IMEC, KUL, CSL; France: CEA, OAMP; Germany: MPE, MPIA; Italy: IFSI, OAP/OAT, OAA/CAISMI, LENS, SISSA; Spain: IAC.	
\end{acknowledgements}

\addcontentsline{toc}{chapter}{Bibliography}
\bibliographystyle{aa}
\bibliography{17635.bib}

\begin{thebibliography}{90}
\expandafter\ifx\csname natexlab\endcsname\relax\def\natexlab#1{#1}\fi

\bibitem[{{Ag{\'u}ndez} {et~al.}(2010){Ag{\'u}ndez}, {Cernicharo},
  {Gu{\'e}lin}, {Kahane}, {Roueff}, {K{\l}os}, {Aoiz}, {Lique}, {Marcelino},
  {Goicoechea}, {Gonz{\'a}lez Garc{\'{\i}}a}, {Gottlieb}, {McCarthy}, \&
  {Thaddeus}}]{agundez2010_cnmin}
{Ag{\'u}ndez}, M., {Cernicharo}, J., {Gu{\'e}lin}, M., {et~al.} 2010, \aap,
  517, L2

\bibitem[{{Begemann} {et~al.}(1994){Begemann}, {Mutschke}, {Dorschner}, \&
  {Henning}}]{begemann1994}
{Begemann}, B., {Mutschke}, H., {Dorschner}, J., \& {Henning}, T. 1994, in
  American Institute of Physics Conference Series, Vol. 312, Molecules and
  Grains in Space, ed. {I.~Nenner}, 781

\bibitem[{{Bohren} \& {Huffman}(1983)}]{bohren1983}
{Bohren}, C.~F. \& {Huffman}, D.~R. 1983, {Absorption and scattering of light
  by small particles} (Wiley)

\bibitem[{{Brown} \& {Millar}(2003)}]{brown2003}
{Brown}, J.~M. \& {Millar}, T.~J. 2003, \mnras, 339, 1041

\bibitem[{{Cernicharo} {et~al.}(2011{\natexlab{a}}){Cernicharo}, {Ag{\'u}ndez},
  {Kahane}, {Gu{\'e}lin}, {Goicoechea}, {Marcelino}, {De Beck}, \&
  {Decin}}]{cernicharo2011_hcnvib}
{Cernicharo}, J., {Ag{\'u}ndez}, M., {Kahane}, C., {et~al.} 2011{\natexlab{a}},
  \aap, 529, L3

\bibitem[{{Cernicharo} {et~al.}(2010{\natexlab{a}}){Cernicharo}, {Decin},
  {Barlow}, {Ag{\'u}ndez}, {Royer}, {Vandenbussche}, {Wesson}, {Polehampton},
  {De Beck}, {Blommaert}, {Daniel}, {De Meester}, {Exter}, {Feuchtgruber},
  {Gear}, {Goicoechea}, {Gomez}, {Groenewegen}, {Hargrave}, {Huygen}, {Imhof},
  {Ivison}, {Jean}, {Kerschbaum}, {Leeks}, {Lim}, {Matsuura}, {Olofsson},
  {Posch}, {Regibo}, {Savini}, {Sibthorpe}, {Swinyard}, {Vandenbussche}, \&
  {Waelkens}}]{cernicharo2010_hcl}
{Cernicharo}, J., {Decin}, L., {Barlow}, M.~J., {et~al.} 2010{\natexlab{a}},
  \aap, 518, L136

\bibitem[{{Cernicharo} \& {Gu\'elin}(1996)}]{cernicharo1996_c8hdetection}
{Cernicharo}, J. \& {Gu\'elin}, M. 1996, \aap, 309, L27

\bibitem[{{Cernicharo} {et~al.}(2007){Cernicharo}, {Gu{\'e}lin}, {Ag{\'u}ndez},
  {Kawaguchi}, {McCarthy}, \& {Thaddeus}}]{cernicharo2007}
{Cernicharo}, J., {Gu{\'e}lin}, M., {Ag{\'u}ndez}, M., {et~al.} 2007, \aap,
  467, L37

\bibitem[{{Cernicharo} {et~al.}(2008){Cernicharo}, {Gu{\'e}lin}, {Ag{\'u}ndez},
  {McCarthy}, \& {Thaddeus}}]{cernicharo2008}
{Cernicharo}, J., {Gu{\'e}lin}, M., {Ag{\'u}ndez}, M., {McCarthy}, M.~C., \&
  {Thaddeus}, P. 2008, \apjl, 688, L83

\bibitem[{{Cernicharo} {et~al.}(2000){Cernicharo}, {Gu{\'e}lin}, \&
  {Kahane}}]{cernicharo2000_2mm}
{Cernicharo}, J., {Gu{\'e}lin}, M., \& {Kahane}, C. 2000, \aaps, 142, 181

\bibitem[{{Cernicharo} {et~al.}(1987{\natexlab{a}}){Cernicharo}, {Gu\'elin},
  {Menten}, \& {Walmsley}}]{cernicharo1987_c6hhyperfine}
{Cernicharo}, J., {Gu\'elin}, M., {Menten}, K.~M., \& {Walmsley}, C.~M.
  1987{\natexlab{a}}, \aap, 181, L1

\bibitem[{{Cernicharo} {et~al.}(1987{\natexlab{b}}){Cernicharo}, {Gu\'elin}, \&
  {Walmsley}}]{cernicharo1987_c5hhyperfine}
{Cernicharo}, J., {Gu\'elin}, M., \& {Walmsley}, C.~M. 1987{\natexlab{b}},
  \aap, 172, L5

\bibitem[{{Cernicharo} {et~al.}(1986{\natexlab{a}}){Cernicharo}, {Kahane},
  {Gomez-Gonzalez}, \& {Gu\'elin}}]{cernicharo1986_c5hdetection}
{Cernicharo}, J., {Kahane}, C., {Gomez-Gonzalez}, J., \& {Gu\'elin}, M.
  1986{\natexlab{a}}, \aap, 167, L5

\bibitem[{{Cernicharo} {et~al.}(1986{\natexlab{b}}){Cernicharo}, {Kahane},
  {Gomez-Gonzalez}, \& {Gu\'elin}}]{cernicharo1986_c5htentativedetection}
---. 1986{\natexlab{b}}, \aap, 164, L1

\bibitem[{{Cernicharo} {et~al.}(2011{\natexlab{b}}){Cernicharo}, {Waters},
  {Decin}, {Encrenaz}, {Tielens}, {Ag\'undez}, \& {De
  Beck}}]{cernicharo2011_scan}
{Cernicharo}, J., {Waters}, L. B. F.~M., {Decin}, L., {et~al.}
  2011{\natexlab{b}}, \emph{in prep.}

\bibitem[{{Cernicharo} {et~al.}(2010{\natexlab{b}}){Cernicharo}, {Waters},
  {Decin}, {Encrenaz}, {Tielens}, {Ag{\'u}ndez}, {De Beck}, {M{\"u}ller},
  {Goicoechea}, {Barlow}, {Benz}, {Crimier}, {Daniel}, {di Giorgio}, {Fich},
  {Gaier}, {Garc{\'{\i}}a-Lario}, {de Koter}, {Khouri}, {Liseau}, {Lombaert},
  {Erickson}, {Pardo}, {Pearson}, {Shipman}, {S{\'a}nchez Contreras}, \&
  {Teyssier}}]{cernicharo2010_hifi1b}
{Cernicharo}, J., {Waters}, L.~B.~F.~M., {Decin}, L., {et~al.}
  2010{\natexlab{b}}, \aap, 521, L8

\bibitem[{{Cernicharo} {et~al.}(1999){Cernicharo}, {Yamamura},
  {Gonz{\'a}lez-Alfonso}, {de Jong}, {Heras}, {Escribano}, \&
  {Ortigoso}}]{cernicharo1999_iso}
{Cernicharo}, J., {Yamamura}, I., {Gonz{\'a}lez-Alfonso}, E., {et~al.} 1999,
  \apjl, 526, L41

\bibitem[{{{Cernicharo}, J. and Kahane} {et~al.}(2011){{Cernicharo}, J. and
  Kahane}, {Gu\'elin}, \& {Ag\'undez}}]{cernicharo2011_3mm}
{{Cernicharo}, J. and Kahane}, M., {Gu\'elin}, M., \& {Ag\'undez}, M. 2011,
  \textit{in prep.}

\bibitem[{{Cordiner} \& {Millar}(2009)}]{cordiner2009}
{Cordiner}, M.~A. \& {Millar}, T.~J. 2009, \apj, 697, 68

\bibitem[{{Crosas} \& {Menten}(1997)}]{crosas1997}
{Crosas}, M. \& {Menten}, K.~M. 1997, \apj, 483, 913

\bibitem[{{De Beck} {et~al.}(2010){De Beck}, {Decin}, {de Koter}, {Justtanont},
  {Verhoelst}, {Kemper}, \& {Menten}}]{debeck2010}
{De Beck}, E., {Decin}, L., {de Koter}, A., {et~al.} 2010, \aap, 523, A18

\bibitem[{{de Graauw} {et~al.}(2010){de Graauw}, {Helmich}, {Phillips},
  {Stutzki}, {Caux}, {Whyborn}, {Dieleman}, {Roelfsema}, {Aarts}, {Assendorp},
  {Bachiller}, {Baechtold}, {Barcia}, {Beintema}, {Belitsky}, {Benz}, {Bieber},
  {Boogert}, {Borys}, {Bumble}, {Ca{\"i}s}, {Caris}, {Cerulli-Irelli},
  {Chattopadhyay}, {Cherednichenko}, {Ciechanowicz}, {Coeur-Joly}, {Comito},
  {Cros}, {de Jonge}, {de Lange}, {Delforges}, {Delorme}, {den Boggende},
  {Desbat}, {Diez-Gonz{\'a}lez}, {di Giorgio}, {Dubbeldam}, {Edwards},
  {Eggens}, {Erickson}, {Evers}, {Fich}, {Finn}, {Franke}, {Gaier}, {Gal},
  {Gao}, {Gallego}, {Gauffre}, {Gill}, {Glenz}, {Golstein}, {Goulooze},
  {Gunsing}, {G{\"u}sten}, {Hartogh}, {Hatch}, {Higgins}, {Honingh}, {Huisman},
  {Jackson}, {Jacobs}, {Jacobs}, {Jarchow}, {Javadi}, {Jellema}, {Justen},
  {Karpov}, {Kasemann}, {Kawamura}, {Keizer}, {Kester}, {Klapwijk}, {Klein},
  {Kollberg}, {Kooi}, {Kooiman}, {Kopf}, {Krause}, {Krieg}, {Kramer},
  {Kruizenga}, {Kuhn}, {Laauwen}, {Lai}, {Larsson}, {Leduc}, {Leinz}, {Lin},
  {Liseau}, {Liu}, {Loose}, {L{\'o}pez-Fernandez}, {Lord}, {Luinge}, {Marston},
  {Mart{\'{\i}}n-Pintado}, {Maestrini}, {Maiwald}, {McCoey}, {Mehdi}, {Megej},
  {Melchior}, {Meinsma}, {Merkel}, {Michalska}, {Monstein}, {Moratschke},
  {Morris}, {Muller}, {Murphy}, {Naber}, {Natale}, {Nowosielski}, {Nuzzolo},
  {Olberg}, {Olbrich}, {Orfei}, {Orleanski}, {Ossenkopf}, {Peacock}, {Pearson},
  {Peron}, {Phillip-May}, {Piazzo}, {Planesas}, {Rataj}, {Ravera}, {Risacher},
  {Salez}, {Samoska}, {Saraceno}, {Schieder}, {Schlecht}, {Schl{\"o}der},
  {Schm{\"u}lling}, {Schultz}, {Schuster}, {Siebertz}, {Smit}, {Szczerba},
  {Shipman}, {Steinmetz}, {Stern}, {Stokroos}, {Teipen}, {Teyssier}, {Tils},
  {Trappe}, {van Baaren}, {van Leeuwen}, {van de Stadt}, {Visser}, {Wildeman},
  {Wafelbakker}, {Ward}, {Wesselius}, {Wild}, {Wulff}, {Wunsch}, {Tielens},
  {Zaal}, {Zirath}, {Zmuidzinas}, \& {Zwart}}]{degraauw2010}
{de Graauw}, T., {Helmich}, F.~P., {Phillips}, T.~G., {et~al.} 2010, \aap, 518,
  L6

\bibitem[{{Decin} {et~al.}(2010{\natexlab{a}}){Decin}, {Ag{\'u}ndez}, {Barlow},
  {Daniel}, {Cernicharo}, {Lombaert}, {De Beck}, {Royer}, {Vandenbussche},
  {Wesson}, {Polehampton}, {Blommaert}, {De Meester}, {Exter}, {Feuchtgruber},
  {Gear}, {Gomez}, {Groenewegen}, {Gu{\'e}lin}, {Hargrave}, {Huygen}, {Imhof},
  {Ivison}, {Jean}, {Kahane}, {Kerschbaum}, {Leeks}, {Lim}, {Matsuura},
  {Olofsson}, {Posch}, {Regibo}, {Savini}, {Sibthorpe}, {Swinyard}, {Yates}, \&
  {Waelkens}}]{decin2010_water}
{Decin}, L., {Ag{\'u}ndez}, M., {Barlow}, M.~J., {et~al.} 2010{\natexlab{a}},
  \nat, 467, 64

\bibitem[{{Decin} {et~al.}(2010{\natexlab{b}}){Decin}, {Cernicharo}, {Barlow},
  {Royer}, {Vandenbussche}, {Wesson}, {Polehampton}, {De Beck}, {Ag{\'u}ndez},
  {Blommaert}, {Cohen}, {Daniel}, {De Meester}, {Exter}, {Feuchtgruber},
  {Fonfria}, {Gear}, {Goicoechea}, {Gomez}, {Groenewegen}, {Hargrave},
  {Huygen}, {Imhof}, {Ivison}, {Jean}, {Kerschbaum}, {Leeks}, {Lim},
  {Matsuura}, {Olofsson}, {Posch}, {Regibo}, {Savini}, {Sibthorpe}, {Swinyard},
  {Tercero}, {Waelkens}, {Witherick}, \& {Yates}}]{decin2010_silicon}
{Decin}, L., {Cernicharo}, J., {Barlow}, M.~J., {et~al.} 2010{\natexlab{b}},
  ArXiv e-prints

\bibitem[{{Decin} {et~al.}(2010{\natexlab{c}}){Decin}, {De Beck},
  {Br{\"u}nken}, {M{\"u}ller}, {Menten}, {Kim}, {Willacy}, {de Koter}, \&
  {Wyrowski}}]{decin2010_gastronoom}
{Decin}, L., {De Beck}, E., {Br{\"u}nken}, S., {et~al.} 2010{\natexlab{c}},
  \aap, 516, A69

\bibitem[{{Decin} {et~al.}(2006){Decin}, {Hony}, {de Koter}, {Justtanont},
  {Tielens}, \& {Waters}}]{decin2006_gastronoom}
{Decin}, L., {Hony}, S., {de Koter}, A., {et~al.} 2006, \aap, 456, 549

\bibitem[{{Decin} {et~al.}(2011){Decin}, {Royer}, {Cox}, {Vandenbussche},
  {Ottensamer}, {Blommaert}, {Groenewegen}, {Barlow}, {Lim}, {Kerschbaum},
  {Posch}, \& {Waelkens}}]{decin2011_shells}
{Decin}, L., {Royer}, P., {Cox}, N.~L.~J., {et~al.} 2011, \aap, 534, A1

\bibitem[{{Deguchi} \& {Nguyen-Q-Rieu}(1990)}]{deguchi1990}
{Deguchi}, S. \& {Nguyen-Q-Rieu}. 1990, \apjl, 360, L27

\bibitem[{{Deleon} \& {Muenter}(1984)}]{deleon1984}
{Deleon}, R.~L. \& {Muenter}, J.~S. 1984, \jcp, 80, 3992

\bibitem[{{Dinh-V-Trung} \& {Lim}(2008)}]{dinh-v-trung2008}
{Dinh-V-Trung} \& {Lim}, J. 2008, \apj, 678, 303

\bibitem[{{Draine}(1978)}]{draine1978}
{Draine}, B.~T. 1978, \apjs, 36, 595

\bibitem[{{Dumouchel} {et~al.}(2010){Dumouchel}, {Faure}, \&
  {Lique}}]{dumouchel2010}
{Dumouchel}, F., {Faure}, A., \& {Lique}, F. 2010, \mnras, 406, 2488

\bibitem[{{Fonfr{\'{\i}}a} {et~al.}(2008){Fonfr{\'{\i}}a}, {Cernicharo},
  {Richter}, \& {Lacy}}]{fonfria2008}
{Fonfr{\'{\i}}a}, J.~P., {Cernicharo}, J., {Richter}, M.~J., \& {Lacy}, J.~H.
  2008, \apj, 673, 445

\bibitem[{{Fong} {et~al.}(2003){Fong}, {Meixner}, \& {Shah}}]{fong2003}
{Fong}, D., {Meixner}, M., \& {Shah}, R.~Y. 2003, \apjl, 582, L39

\bibitem[{{Gonz{\'a}lez-Alfonso} \& {Cernicharo}(1999)}]{gonzalez-alfonso1999}
{Gonz{\'a}lez-Alfonso}, E. \& {Cernicharo}, J. 1999, in ESA Special
  Publication, Vol. 427, The Universe as Seen by ISO, ed. {P.~Cox \&
  M.~Kessler}, 325

\bibitem[{{Goorvitch} \& {Chackerian}(1994)}]{goorvitch1994}
{Goorvitch}, D. \& {Chackerian}, Jr., C. 1994, \apjs, 91, 483

\bibitem[{{Griffin} {et~al.}(2010){Griffin}, {Abergel}, {Abreu}, {Ade},
  {Andr{\'e}}, {Augueres}, {Babbedge}, {Bae}, {Baillie}, {Baluteau}, {Barlow},
  {Bendo}, {Benielli}, {Bock}, {Bonhomme}, {Brisbin}, {Brockley-Blatt},
  {Caldwell}, {Cara}, {Castro-Rodriguez}, {Cerulli}, {Chanial}, {Chen},
  {Clark}, {Clements}, {Clerc}, {Coker}, {Communal}, {Conversi}, {Cox},
  {Crumb}, {Cunningham}, {Daly}, {Davis}, {de Antoni}, {Delderfield}, {Devin},
  {di Giorgio}, {Didschuns}, {Dohlen}, {Donati}, {Dowell}, {Dowell}, {Duband},
  {Dumaye}, {Emery}, {Ferlet}, {Ferrand}, {Fontignie}, {Fox}, {Franceschini},
  {Frerking}, {Fulton}, {Garcia}, {Gastaud}, {Gear}, {Glenn}, {Goizel},
  {Griffin}, {Grundy}, {Guest}, {Guillemet}, {Hargrave}, {Harwit}, {Hastings},
  {Hatziminaoglou}, {Herman}, {Hinde}, {Hristov}, {Huang}, {Imhof}, {Isaak},
  {Israelsson}, {Ivison}, {Jennings}, {Kiernan}, {King}, {Lange}, {Latter},
  {Laurent}, {Laurent}, {Leeks}, {Lellouch}, {Levenson}, {Li}, {Li},
  {Lilienthal}, {Lim}, {Liu}, {Lu}, {Madden}, {Mainetti}, {Marliani}, {McKay},
  {Mercier}, {Molinari}, {Morris}, {Moseley}, {Mulder}, {Mur}, {Naylor},
  {Nguyen}, {O'Halloran}, {Oliver}, {Olofsson}, {Olofsson}, {Orfei}, {Page},
  {Pain}, {Panuzzo}, {Papageorgiou}, {Parks}, {Parr-Burman}, {Pearce},
  {Pearson}, {P{\'e}rez-Fournon}, {Pinsard}, {Pisano}, {Podosek}, {Pohlen},
  {Polehampton}, {Pouliquen}, {Rigopoulou}, {Rizzo}, {Roseboom}, {Roussel},
  {Rowan-Robinson}, {Rownd}, {Saraceno}, {Sauvage}, {Savage}, {Savini},
  {Sawyer}, {Scharmberg}, {Schmitt}, {Schneider}, {Schulz}, {Schwartz},
  {Shafer}, {Shupe}, {Sibthorpe}, {Sidher}, {Smith}, {Smith}, {Smith},
  {Spencer}, {Stobie}, {Sudiwala}, {Sukhatme}, {Surace}, {Stevens}, {Swinyard},
  {Trichas}, {Tourette}, {Triou}, {Tseng}, {Tucker}, {Turner}, {Vaccari},
  {Valtchanov}, {Vigroux}, {Virique}, {Voellmer}, {Walker}, {Ward}, {Waskett},
  {Weilert}, {Wesson}, {White}, {Whitehouse}, {Wilson}, {Winter}, {Woodcraft},
  {Wright}, {Xu}, {Zavagno}, {Zemcov}, {Zhang}, \& {Zonca}}]{griffin2010}
{Griffin}, M.~J., {Abergel}, A., {Abreu}, A., {et~al.} 2010, \aap, 518, L3

\bibitem[{{Groenewegen} {et~al.}(1996){Groenewegen}, {Baas}, {de Jong}, \&
  {Loup}}]{groenewegen1996_co}
{Groenewegen}, M.~A.~T., {Baas}, F., {de Jong}, T., \& {Loup}, C. 1996, \aap,
  306, 241

\bibitem[{{Groenewegen} {et~al.}(1998){Groenewegen}, {van der Veen}, \&
  {Matthews}}]{groenewegen1998}
{Groenewegen}, M.~A.~T., {van der Veen}, W.~E.~C.~J., \& {Matthews}, H.~E.
  1998, \aap, 338, 491

\bibitem[{{Groenewegen} {et~al.}(2011){Groenewegen}, {Waelkens}, {Barlow},
  {Kerschbaum}, {Garcia-Lario}, {Cernicharo}, {Blommaert}, {Bouwman}, {Cohen},
  {Cox}, {Decin}, {Exter}, {Gear}, {Gomez}, {Hargrave}, {Henning},
  {Hutsem{\'e}kers}, {Ivison}, {Jorissen}, {Krause}, {Ladjal}, {Leeks}, {Lim},
  {Matsuura}, {Naz{\'e}}, {Olofsson}, {Ottensamer}, {Polehampton}, {Posch},
  {Rauw}, {Royer}, {Sibthorpe}, {Swinyard}, {Ueta}, {Vamvatira-Nakou},
  {Vandenbussche}, {van de Steene}, {van Eck}, {van Hoof}, {van Winckel},
  {Verdugo}, \& {Wesson}}]{groenewegen2011}
{Groenewegen}, M.~A.~T., {Waelkens}, C., {Barlow}, M.~J., {et~al.} 2011, \aap,
  526, A162

\bibitem[{{Gu\'elin} {et~al.}(1987){Gu\'elin}, {Cernicharo}, {Kahane},
  {Gomez-Gonzalez}, \& {Walmsley}}]{guelin1987_c6hdetection}
{Gu\'elin}, M., {Cernicharo}, J., {Kahane}, C., {Gomez-Gonzalez}, J., \&
  {Walmsley}, C.~M. 1987, \aap, 175, L5

\bibitem[{{Gu\'elin} {et~al.}(1997){Gu\'elin}, {Cernicharo}, {Travers},
  {McCarthy}, {Gottlieb}, {Thaddeus}, {Ohishi}, {Saito}, \&
  {Yamamoto}}]{guelin1997_c7h}
{Gu\'elin}, M., {Cernicharo}, J., {Travers}, M.~J., {et~al.} 1997, \aap, 317,
  L1

\bibitem[{{Gu\'elin} {et~al.}(1978){Gu\'elin}, {Green}, \&
  {Thaddeus}}]{guelin1978_c4hdetection}
{Gu\'elin}, M., {Green}, S., \& {Thaddeus}, P. 1978, \apjl, 224, L27

\bibitem[{{Gu\'elin} {et~al.}(1993){Gu\'elin}, {Lucas}, \&
  {Cernicharo}}]{guelin1993}
{Gu\'elin}, M., {Lucas}, R., \& {Cernicharo}, J. 1993, \aap, 280, L19

\bibitem[{{Gu{\'e}lin} {et~al.}(1999){Gu{\'e}lin}, {Neininger}, {Lucas}, \&
  {Cernicharo}}]{guelin1999}
{Gu{\'e}lin}, M., {Neininger}, N., {Lucas}, R., \& {Cernicharo}, J. 1999, in
  The Physics and Chemistry of the Interstellar Medium, ed. {V.~Ossenkopf,
  J.~Stutzki, \& G.~Winnewisser}, 326

\bibitem[{{Habing} \& {Olofsson}(2003)}]{habing2003}
{Habing}, H.~J. \& {Olofsson}, H., eds. 2003, {Asymptotic giant branch stars}

\bibitem[{{He} {et~al.}(2008){He}, {Dinh-V-Trung}, {Kwok}, {M{\"u}ller},
  {Zhang}, {Hasegawa}, {Peng}, \& {Huang}}]{he2008}
{He}, J.~H., {Dinh-V-Trung}, {Kwok}, S., {et~al.} 2008, \apjs, 177, 275

\bibitem[{{Huggins} {et~al.}(1988){Huggins}, {Olofsson}, \&
  {Johansson}}]{huggins1988_co}
{Huggins}, P.~J., {Olofsson}, H., \& {Johansson}, L.~E.~B. 1988, \apj, 332,
  1009

\bibitem[{{Kahane} {et~al.}(1988){Kahane}, {Gomez-Gonzalez}, {Cernicharo}, \&
  {Gu\'elin}}]{kahane1988}
{Kahane}, C., {Gomez-Gonzalez}, J., {Cernicharo}, J., \& {Gu\'elin}, M. 1988,
  \aap, 190, 167

\bibitem[{{Kahane} {et~al.}(2011){Kahane}, {Cernicharo}, {Gu\'elin}, \&
  {Ag\'undez}}]{kahane2011_1mm}
{Kahane}, M., {Cernicharo}, J., {Gu\'elin}, M., \& {Ag\'undez}, M. 2011,
  \textit{in prep.}

\bibitem[{{Kama} {et~al.}(2009){Kama}, {Min}, \& {Dominik}}]{kama2009}
{Kama}, M., {Min}, M., \& {Dominik}, C. 2009, \aap, 506, 1199

\bibitem[{{Kawaguchi} {et~al.}(2007){Kawaguchi}, {Fujimori}, {Aimi}, {Takano},
  {Okabayashi}, {Gupta}, {Br{\"u}nken}, {Gottlieb}, {McCarthy}, \&
  {Thaddeus}}]{kawaguchi2007}
{Kawaguchi}, K., {Fujimori}, R., {Aimi}, S., {et~al.} 2007, \pasj, 59, L47

\bibitem[{{Kramer}(1997)}]{kramer1997}
{Kramer}, C. 1997, Calibration of spectral line data at the IRAM 30m radio
  telescope, Tech. rep., IRAM

\bibitem[{{Ladjal} {et~al.}(2010){Ladjal}, {Justtanont}, {Groenewegen},
  {Blommaert}, {Waelkens}, \& {Barlow}}]{ladjal2010}
{Ladjal}, D., {Justtanont}, K., {Groenewegen}, M.~A.~T., {et~al.} 2010, \aap,
  513, A53

\bibitem[{{Larsson} {et~al.}(2002){Larsson}, {Liseau}, \&
  {Men'shchikov}}]{larsson2002}
{Larsson}, B., {Liseau}, R., \& {Men'shchikov}, A.~B. 2002, \aap, 386, 1055

\bibitem[{{Le{\~a}o} {et~al.}(2006){Le{\~a}o}, {de Laverny}, {M{\'e}karnia},
  {de Medeiros}, \& {Vandame}}]{leao2006}
{Le{\~a}o}, I.~C., {de Laverny}, P., {M{\'e}karnia}, D., {de Medeiros}, J.~R.,
  \& {Vandame}, B. 2006, \aap, 455, 187

\bibitem[{{Le Bertre}(1992)}]{lebertre1992}
{Le Bertre}, T. 1992, \aaps, 94, 377

\bibitem[{{Lombaert} {et~al.}(2012){Lombaert}, {Decin}, {Blommaert}, {de
  Koter}, {de Vries}, {De Beck}, {Min}, \& {Waters}}]{lombaert2011}
{Lombaert}, R., {Decin}, L., {Blommaert}, J. A. D.~L., {et~al.} 2012,
  \textit{in prep.}

\bibitem[{{Loup} {et~al.}(1993){Loup}, {Forveille}, {Omont}, \&
  {Paul}}]{loup1993}
{Loup}, C., {Forveille}, T., {Omont}, A., \& {Paul}, J.~F. 1993, \aaps, 99, 291

\bibitem[{{Maercker} {et~al.}(2008){Maercker}, {Sch{\"o}ier}, {Olofsson},
  {Bergman}, \& {Ramstedt}}]{maercker2008}
{Maercker}, M., {Sch{\"o}ier}, F.~L., {Olofsson}, H., {Bergman}, P., \&
  {Ramstedt}, S. 2008, \aap, 479, 779

\bibitem[{{Mauron} \& {Huggins}(1999)}]{mauron1999}
{Mauron}, N. \& {Huggins}, P.~J. 1999, \aap, 349, 203

\bibitem[{{Men'shchikov} {et~al.}(2001){Men'shchikov}, {Balega}, {Bl{\"o}cker},
  {Osterbart}, \& {Weigelt}}]{menshchikov2001}
{Men'shchikov}, A.~B., {Balega}, Y., {Bl{\"o}cker}, T., {Osterbart}, R., \&
  {Weigelt}, G. 2001, \aap, 368, 497

\bibitem[{{Min} {et~al.}(2009){Min}, {Dullemond}, {Dominik}, {de Koter}, \&
  {Hovenier}}]{min2009_mcmax}
{Min}, M., {Dullemond}, C.~P., {Dominik}, C., {de Koter}, A., \& {Hovenier},
  J.~W. 2009, \aap, 497, 155

\bibitem[{{Min} {et~al.}(2003){Min}, {Hovenier}, \& {de Koter}}]{min2003_dhs}
{Min}, M., {Hovenier}, J.~W., \& {de Koter}, A. 2003, \aap, 404, 35

\bibitem[{{Monnier} {et~al.}(2000){Monnier}, {Danchi}, {Hale}, {Tuthill}, \&
  {Townes}}]{monnier2000}
{Monnier}, J.~D., {Danchi}, W.~C., {Hale}, D.~S., {Tuthill}, P.~G., \&
  {Townes}, C.~H. 2000, \apj, 543, 868

\bibitem[{{Mordaunt} {et~al.}(1998){Mordaunt}, {Ashfold}, {Dixon},
  {L{\"o}ffler}, {Schnieder}, \& {Welge}}]{mordaunt1998}
{Mordaunt}, D.~H., {Ashfold}, M.~N.~R., {Dixon}, R.~N., {et~al.} 1998, \jcp,
  108, 519

\bibitem[{{M{\"u}ller} {et~al.}(2000){M{\"u}ller}, {Klaus}, \&
  {Winnewisser}}]{mueller2000}
{M{\"u}ller}, H.~S.~P., {Klaus}, T., \& {Winnewisser}, G. 2000, \aap, 357, L65

\bibitem[{{M{\"u}ller} {et~al.}(2005){M{\"u}ller}, {Schl{\"o}der}, {Stutzki},
  \& {Winnewisser}}]{mueller2005_cdms}
{M{\"u}ller}, H.~S.~P., {Schl{\"o}der}, F., {Stutzki}, J., \& {Winnewisser}, G.
  2005, Journal of Molecular Structure, 742, 215

\bibitem[{{Olofsson} {et~al.}(1993){Olofsson}, {Eriksson}, {Gustafsson}, \&
  {Carlstrom}}]{olofsson1993_co}
{Olofsson}, H., {Eriksson}, K., {Gustafsson}, B., \& {Carlstrom}, U. 1993,
  \apjs, 87, 267

\bibitem[{{Padovani} {et~al.}(2009){Padovani}, {Walmsley}, {Tafalla}, {Galli},
  \& {M{\"u}ller}}]{padovani2009}
{Padovani}, M., {Walmsley}, C.~M., {Tafalla}, M., {Galli}, D., \& {M{\"u}ller},
  H.~S.~P. 2009, \aap, 505, 1199

\bibitem[{{Pilbratt} {et~al.}(2010){Pilbratt}, {Riedinger}, {Passvogel},
  {Crone}, {Doyle}, {Gageur}, {Heras}, {Jewell}, {Metcalfe}, {Ott}, \&
  {Schmidt}}]{pilbratt2010}
{Pilbratt}, G.~L., {Riedinger}, J.~R., {Passvogel}, T., {et~al.} 2010, \aap,
  518, L1

\bibitem[{{Pitman} {et~al.}(2008){Pitman}, {Hofmeister}, {Corman}, \&
  {Speck}}]{pitman2008}
{Pitman}, K.~M., {Hofmeister}, A.~M., {Corman}, A.~B., \& {Speck}, A.~K. 2008,
  \aap, 483, 661

\bibitem[{{Poglitsch} {et~al.}(2010){Poglitsch}, {Waelkens}, {Geis},
  {Feuchtgruber}, {Vandenbussche}, {Rodriguez}, {Krause}, {Renotte}, {van
  Hoof}, {Saraceno}, {Cepa}, {Kerschbaum}, {Agn{\`e}se}, {Ali}, {Altieri},
  {Andreani}, {Augueres}, {Balog}, {Barl}, {Bauer}, {Belbachir}, {Benedettini},
  {Billot}, {Boulade}, {Bischof}, {Blommaert}, {Callut}, {Cara}, {Cerulli},
  {Cesarsky}, {Contursi}, {Creten}, {De Meester}, {Doublier}, {Doumayrou},
  {Duband}, {Exter}, {Genzel}, {Gillis}, {Gr{\"o}zinger}, {Henning},
  {Herreros}, {Huygen}, {Inguscio}, {Jakob}, {Jamar}, {Jean}, {de Jong},
  {Katterloher}, {Kiss}, {Klaas}, {Lemke}, {Lutz}, {Madden}, {Marquet},
  {Martignac}, {Mazy}, {Merken}, {Montfort}, {Morbidelli}, {M{\"u}ller},
  {Nielbock}, {Okumura}, {Orfei}, {Ottensamer}, {Pezzuto}, {Popesso},
  {Putzeys}, {Regibo}, {Reveret}, {Royer}, {Sauvage}, {Schreiber}, {Stegmaier},
  {Schmitt}, {Schubert}, {Sturm}, {Thiel}, {Tofani}, {Vavrek}, {Wetzstein},
  {Wieprecht}, \& {Wiezorrek}}]{poglitsch2010}
{Poglitsch}, A., {Waelkens}, C., {Geis}, N., {et~al.} 2010, \aap, 518, L2

\bibitem[{{Preibisch} {et~al.}(1993){Preibisch}, {Ossenkopf}, {Yorke}, \&
  {Henning}}]{preibisch1993}
{Preibisch}, T., {Ossenkopf}, V., {Yorke}, H.~W., \& {Henning}, T. 1993, \aap,
  279, 577

\bibitem[{{Ramstedt} {et~al.}(2008){Ramstedt}, {Sch{\"o}ier}, {Olofsson}, \&
  {Lundgren}}]{ramstedt2008}
{Ramstedt}, S., {Sch{\"o}ier}, F.~L., {Olofsson}, H., \& {Lundgren}, A.~A.
  2008, \aap, 487, 645

\bibitem[{{Remijan} {et~al.}(2007){Remijan}, {Hollis}, {Lovas}, {Cordiner},
  {Millar}, {Markwick-Kemper}, \& {Jewell}}]{remijan2007}
{Remijan}, A.~J., {Hollis}, J.~M., {Lovas}, F.~J., {et~al.} 2007, \apjl, 664,
  L47

\bibitem[{{Ridgway} \& {Keady}(1988)}]{ridgway1988}
{Ridgway}, S. \& {Keady}, J.~J. 1988, \apj, 326, 843

\bibitem[{{Roelfsema} {et~al.}(2012){Roelfsema}, {Helmich}, {Teyssier},
  {Ossenkopf}, {Morris}, {Olberg}, {Shipman}, {Risacher}, {Akyilmaz},
  {Assendorp}, {Avruch}, {Beintema}, {Biver}, {Boogert}, {Borys}, {Braine},
  {Caris}, {Caux}, {Cernicharo}, {Coeur-Joly}, {Comito}, {de Lange},
  {Delforge}, {Dieleman}, {Dubbeldam}, {de Graauw}, {Edwards}, {Fich},
  {Flederus}, {Gal}, {di Giorgio}, {Herpin}, {Higgins}, {Hoac}, {Huisman},
  {Jarchow}, {Jellema}, {de Jonge}, {Kester}, {Klein}, {Kooi}, {Kramer},
  {Laauwen}, {Larsson}, {Leinz}, {Lord}, {Lorenzani}, {Luinge}, {Marston},
  {Mart{\'{\i}}n-Pintado}, {McCoey}, {Melchior}, {Michalska}, {Moreno},
  {M{\"u}ller}, {Nowosielski}, {Okada}, {Orlea{\'n}ski}, {Phillips}, {Pearson},
  {Rabois}, {Ravera}, {Rector}, {Rengel}, {Sagawa}, {Salomons},
  {S{\'a}nchez-Su{\'a}rez}, {Schieder}, {Schl{\"o}der}, {Schm{\"u}lling},
  {Soldati}, {Stutzki}, {Thomas}, {Tielens}, {Vastel}, {Wildeman}, {Xie},
  {Xilouris}, {Wafelbakker}, {Whyborn}, {Zaal}, {Bell}, {Bjerkeli}, {De Beck},
  {Cavali{\'e}}, {Crockett}, {Hily-Blant}, {Kama}, {Kaminski}, {Lefl{\'o}ch},
  {Lombaert}, {de Luca}, {Makai}, {Marseille}, {Nagy}, {Pacheco}, {van der
  Wiel}, {Wang}, \& {Y{\i}ld{\i}z}}]{roelfsema2012}
{Roelfsema}, P.~R., {Helmich}, F.~P., {Teyssier}, D., {et~al.} 2012, \aap, 537,
  A17

\bibitem[{{Schloerb} {et~al.}(1983){Schloerb}, {Irvine}, {Friberg},
  {Hjalmarson}, \& {Hoglund}}]{schloerb1983}
{Schloerb}, F.~P., {Irvine}, W.~M., {Friberg}, P., {Hjalmarson}, A., \&
  {Hoglund}, B. 1983, \apj, 264, 161

\bibitem[{{Sch{\"o}ier} {et~al.}(2007){Sch{\"o}ier}, {Bast}, {Olofsson}, \&
  {Lindqvist}}]{schoeier2007}
{Sch{\"o}ier}, F.~L., {Bast}, J., {Olofsson}, H., \& {Lindqvist}, M. 2007,
  \aap, 473, 871

\bibitem[{{Skinner} {et~al.}(1999){Skinner}, {Justtanont}, {Tielens}, {Betz},
  {Boreiko}, \& {Baas}}]{skinner1999}
{Skinner}, C.~J., {Justtanont}, K., {Tielens}, A.~G.~G.~M., {et~al.} 1999,
  \mnras, 302, 293

\bibitem[{Tarroni \& Carter(2004)}]{tarroni2004}
Tarroni, R. \& Carter, S. 2004in  (Taylor \& Francis Ltd.), 2167--2179,
  european Network Theonet II Meeting, Bologna, Italy, Nov., 2003

\bibitem[{{Tenenbaum} {et~al.}(2010){Tenenbaum}, {Dodd}, {Milam}, {Woolf}, \&
  {Ziurys}}]{tenenbaum2010}
{Tenenbaum}, E.~D., {Dodd}, J.~L., {Milam}, S.~N., {Woolf}, N.~J., \& {Ziurys},
  L.~M. 2010, \apjl, 720, L102

\bibitem[{{Teyssier} {et~al.}(2006){Teyssier}, {Hernandez}, {Bujarrabal},
  {Yoshida}, \& {Phillips}}]{teyssier2006}
{Teyssier}, D., {Hernandez}, R., {Bujarrabal}, V., {Yoshida}, H., \&
  {Phillips}, T.~G. 2006, \aap, 450, 167

\bibitem[{{Thaddeus} {et~al.}(2008){Thaddeus}, {Gottlieb}, {Gupta},
  {Br{\"u}nken}, {McCarthy}, {Ag{\'u}ndez}, {Gu{\'e}lin}, \&
  {Cernicharo}}]{thaddeus2008}
{Thaddeus}, P., {Gottlieb}, C.~A., {Gupta}, H., {et~al.} 2008, \apj, 677, 1132

\bibitem[{{Tucker} {et~al.}(1974){Tucker}, {Kutner}, \&
  {Thaddeus}}]{tucker1974}
{Tucker}, K.~D., {Kutner}, M.~L., \& {Thaddeus}, P. 1974, \apjl, 193, L115

\bibitem[{{van Dishoeck} {et~al.}(2006){van Dishoeck}, {Jonkheid}, \& {van
  Hemert}}]{vandishoeck2006}
{van Dishoeck}, E.~F., {Jonkheid}, B., \& {van Hemert}, M.~C. 2006, Faraday
  Discussions, 133, 231

\bibitem[{{Wang} {et~al.}(1994){Wang}, {Jaffe}, {Graf}, \& {Evans}}]{wang1994}
{Wang}, Y., {Jaffe}, D.~T., {Graf}, U.~U., \& {Evans}, II, N.~J. 1994, \apjs,
  95, 503

\bibitem[{{Winnewisser} {et~al.}(1997){Winnewisser}, {Belov}, {Klaus}, \&
  {Schieder}}]{winnewisser1997}
{Winnewisser}, G., {Belov}, S.~P., {Klaus}, T., \& {Schieder}, R. 1997, J. Mol.
  Spec., 184, 468

\bibitem[{Woon(1995)}]{woon1995}
Woon, D.~E. 1995, Chemical Physics Letters, 244, 45

\end{thebibliography}

\end{document}